\documentclass[a4paper,fleqn,usenatbib]{mnras}

\usepackage{newtxtext,newtxmath}

\usepackage[T1]{fontenc}
\usepackage{ae,aecompl}

\usepackage{graphicx}	
\usepackage{amsmath}	
\usepackage{amssymb}	

\def\arcsec{$^{\prime\prime}$}

\def\lir{$L_{\rm IR}$}

\def\l1.4{$L_{\rm 1.4GHz}$} \def\s1.4{$S_{\rm 1.4GHz}$}

\def\kms{km\,s$^{-1}$}

\def\gs{\gtrsim}
\def\ls{\lesssim}

\title[The most distant, luminous, dusty, star-forming galaxies]
{The most distant, luminous, dusty star-forming galaxies:\\
         redshifts from NOEMA and ALMA spectral scans}

\author[Fudamoto et al.]{Yoshinobu~Fudamoto,$^{\! 1,2}$\thanks{yoshinobu.fudamoto@unige.ch}
R.\,J.~Ivison,$^{\! 2,3}$
I.~Oteo,$^{\! 3,2}$
M.~Krips,$^{\! 4}$
Z.-Y.~Zhang,$^{\! 3,2}$
A.~Weiss,$^{\! 5}$\newauthor
H.~Dannerbauer,$^{\! 6, 7}$
A.~Omont,$^{\! 8,9}$
S.\,C.~Chapman,$^{\! 10}$
L.~Christensen,$^{\! 11}$
V.~Arumugam,$^{\! 2, 3}$\newauthor
F.~Bertoldi,$^{\! 12}$
M.~Bremer,$^{\! 13}$
D.\,L.~Clements,$^{\! 14}$
L.~Dunne,$^{\! 3,15}$
S.\,A.~Eales,$^{\! 15}$
J.~Greenslade,$^{\! 14}$\newauthor
S.~Maddox,$^{\! 3,15}$
P.~Martinez-Navajas,$^{\! 7}$
M.~Michalowski,$^{\! 3}$
I.~P\'erez-Fournon,$^{\! 6,7}$\newauthor
D.~Riechers,$^{\! 16}$
J.\,M.~Simpson,$^{\! 3,17}$
B.~Stalder,$^{\! 18}$
E.~Valiante$^{15}$ and
P.~van~der~Werf$^{19}$
\vspace*{1mm}\\
$^{1}$ Observatoire de Gen\`{e}ve, 51 Ch.\ des Maillettes, 1290 Versoix, Switzerland\\
$^{2}$ ESO, Karl-Schwarzschild-Str.~2, D-85748 Garching, Germany\\
$^{3}$ Institute for Astronomy, University of Edinburgh, Royal Observatory, Blackford Hill, Edinburgh EH9 3HJ\\
$^{4}$ Institut de RadioAstronomie Millim\'etrique, 300 rue de la Piscine, Domaine Universitaire, 38406 Saint Martin d'H\`eres, France\\
$^{5}$ Max-Planck-Institut f\"{u}r Radioastronomie, Auf dem H\"{u}gel 69, D-53121 Bonn, Germany\\
$^{6}$ IAC, E-38200 La  Laguna, Tenerife, Spain\\
$^{7}$ Departamento de Astrofisica, Universidad de La Laguna, E-38205 La Laguna, Tenerife, Spain\\
$^{8}$ UPMC Univ Paris 06, UMR 7095, IAP, 75014, Paris, France\\
$^{9}$ CNRS, UMR7095, IAP, F-75014, Paris, France\\
$^{10}$ Department of Physics and Atmospheric Science, Dalhousie University, Halifax, NS B3H 4R2, Canada\\
$^{11}$ Dark Cosmology Centre, Niels Bohr Institute, University of Copenhagen, Juliane Maries Vej 30, DK-2100 Copenhagen, Denmark\\
$^{12}$ Argelander-Institute for Astronomy, Bonn University, Auf dem Huegel 71, 53121 Bonn, Germany\\
$^{13}$ H.\,H.\ Wills Physics Laboratory, University of Bristol, Tyndall Avenue, Bristol BS8 1TL\\
$^{14}$ Astrophysics Group, Imperial College London, Blackett Laboratory, Prince Consort Road, London SW7 2AZ\\
$^{15}$ School of Physics \& Astronomy, Cardiff University, Queen's Buildings, The Parade, Cardiff CF24 3AA\\
$^{16}$ Astronomy Department, Cornell University, Ithaca, NY 14853, USA\\
$^{17}$ Academia Sinica Institute of Astronomy and Astrophysics, No.~1, Sec.~4, Roosevelt Rd, Taipei 10617, Taiwan\\
$^{18}$ Institute for Astronomy, University of Hawaii, 2680 Woodlawn Drive, Honolulu, HI 96822, USA\\
$^{19}$ Leiden Observatory, Leiden University, P.O.\ Box 9513, NL-2300 RA Leiden, The Netherlands
}

\date{Accepted 2017 July 26. Received 2017 July 26; in original form 2017 June 15}

\pubyear{2017}

\begin{document}
\label{firstpage}
\pagerange{\pageref{firstpage}--\pageref{lastpage}}
\maketitle

\begin{abstract}
  We present 1.3- and/or 3-mm continuum images and 3-mm spectral
  scans, obtained using NOEMA and ALMA, of 21 distant, dusty,
  star-forming galaxies (DSFGs). Our sample is a subset of the
  galaxies selected by \citet{ivison16} on the basis of their
  extremely red far-infrared (far-IR) colours and low {\it
    Herschel} flux densities; most are thus expected to be
  unlensed, extraordinarily luminous starbursts at $z\gs 4$,
  modulo the considerable cross-section to gravitational lensing
  implied by their redshift. We observed 17 of these galaxies
  with NOEMA and four with ALMA, scanning through the 3-mm
  atmospheric window.  We have obtained secure redshifts for
  seven galaxies via detection of multiple CO lines, one of them
  a lensed system at $z=6.027$ (two others are also found to be
  lensed); a single emission line was detected in another four
  galaxies, one of which has been shown elsewhere to lie at
  $z=4.002$.  Where we find no spectroscopic redshifts, the
  galaxies are generally less luminous by 0.3--0.4\,dex, which
  goes some way to explaining our failure to detect line
  emission. We show that this sample contains amongst the most
  luminous known star-forming galaxies.  Due to their extreme
  star-formation activity, these galaxies will consume their
  molecular gas in $\ls 100$\,Myr, despite their high molecular
  gas masses, and are therefore plausible progenitors of the
  massive, `red-and-dead' elliptical galaxies at $z \approx 3$.
\end{abstract}


\section{Introduction}

It has been known since the 1970s and 1980s that a large fraction
of the energy produced by vigorously star-forming galaxies in the
nearby Universe is radiated by cool dust which mingles with their
reservoirs of molecular gas \citep*[e.g.][]{snh87}.  A decade on,
the existence of a more distant population of dusty galaxies was
inferred by \citet{puget96} from the detection of the cosmic
far-IR background using FIRAS aboard the {\it Cosmic Background
  Explorer}, individual examples of which were quickly detected
by \citet*{smail97} in the submillimeter (submm) waveband.  If
their initial stellar mass function (IMF) is normal, these
galaxies form stars at tremendous rates, sometimes
$>1000\,{\rm M_{\odot}\,{\rm yr^{-1}}}$ \citep[e.g.][]{ivison98}.
Deeper submm observations in cosmological deep fields
\citep[e.g.][]{barger98, hughes98, eales99} confirmed the
abundance of these so-called submm galaxies (SMGs), sometimes
known now as dusty, star-forming galaxies \citep*[DSFGs ---
e.g.][]{casey14}.

In the decades since then, the SPIRE camera \citep{griffin10}
aboard {\it Herschel} \citep{pilbratt10} and the SCUBA-2 camera
\citep{holland13} on the James Clerk Maxwell Telescope (JCMT)
have together detected orders of magnitude more of these DSFGs.
Conventional optical and near-IR spectroscopic observations
confirmed that DSFGs are considerably more abundant
($\approx 1,000\times$) at high redshift than in the local
Universe, with a redshift distribution for those selected at
850\,$\mu$m that peaks at $z\sim1$--3
\citep[e.g.][]{chapman05,simpson14}.  Those selected at $>1$\,mm
by the South Pole Telescope \citep[e.g.][]{vieira10,strandet17}
are more distant while those selected at the far-IR wavelengths
imaged by {\it Herschel} are typically at $z<2$.

In the local Universe, massive early-type galaxies have old
stellar populations, $>2$\,Gyr, and are therefore red in
optical color -- so-called `red-and-dead' galaxies.  They have
little gas or dust, and star-formation activity has ceased
\citep[see][for a review, cf.\ \citealt{eales17}]{renzini06}.
The majority of these galaxies experienced an intense phase of
star formation around 5--10\,Gyr ago \citep[e.g.][]{thomas10},
and current observational evidence suggests that DSFGs at
$z\approx 2$ are their likely progenitors.

It is also well established that there exists a population of massive
elliptical galaxies at $z\sim 2$--3.  It has been claimed that most of
these are high-redshift analogs of local, massive red-and-dead
galaxies \citep[i.e.\ high stellar masses, red colors, old stellar
populations --- see e.g.][see also \citealt{dunlop96} for a rarer but
similarly old galaxy at $z= 1.55$]{cimatti04, trujillo06, kriek08,
  dokkum10}.  The existence of these galaxies at $z\sim2$--3 suggests
intense star-formation episodes must occur at even higher redshifts,
perhaps implying that DSFGs are common at $z\gs 4$
\citep[e.g.][]{toft14}.

Only a small number of DSFGs were known at $z\gs 4$ until
recently, most of them gravitationally lensed
\citep[e.g.][]{asboth16}.  To address this issue,
\citet{ivison16} recently exploited the widest available far-IR
imaging survey, {\it H}-ATLAS \citep{eales10}, to create a sample
of the faintest, reddest dusty galaxies, further improving their
photometric redshifts via ground-based photometry from SCUBA-2
\citep{holland13} and LABOCA \citep{siringo09}.  The galaxies
thus selected are expected to be largely
unlensed\footnote{Despite expecting a low lensing fraction,
  \citeauthor{ivison16} and others have shown that strongly
  lensed galaxies are common at $z>4$ due to the increase with
  redshift of the optical depth to lensing and the magnification
  bias; \citet{oteo17morph} present high-resolution ALMA imaging
  of this sample, showing that the fraction of lensed galaxies is
  indeed relatively high.}, luminous and very distant.  Their
vigorous star-formation activity thus tallies with the
star-formation history required to build up the large mass of
stars found in spheroidal galaxies at $z\approx 2$.

To confirm that the ultrared DSFGs selected by \citeauthor{ivison16}
do lie at $z\gs 4$, which will strengthen their links with
red-and-dead galaxies at $z\sim2-3$, requires robust spectroscopic
confirmation of their photometric redshifts.  This is non-trivial when
working in the traditional optical and near-IR regime, verging on
impossible with current telescopes and instrumentation.  Following the
success of \citet{cox11}, who scanned the 3-mm atmospheric window to
determine the redshift of one of the brightest, reddest, lensed
galaxies to emerge from {\it H}-ATLAS \citep[see also][]{weiss13}, we
have therefore obtained 3-mm spectral scans of 21 ultrared DSFGs from
the \citeauthor{ivison16} sample, as well as interferometric 0.85- and
1.3-mm imaging to better pinpoint their positions.

Our primary objective here is to determine robust spectroscopic
redshifts for these DSFGs, via the detection of multiple molecular
and/or atomic emission lines.  Using these to fine-tune the
far-IR/submm photometric techniques employed by \citeauthor{ivison16}
then allows us to more reliably determine the space density of DSFGs
at $z\gs 4$.  In addition, we use our improved measurements of IR
luminosity and our CO line luminosities to estimate physical
properties, such as SFR and molecular gas mass.  Finally, we compare
these derived properties with those of other DSFGs at low and high
redshifts, subject as usual to the considerable uncertainties imposed
by $\alpha_{\rm CO}$ and the assumed IMF.

Where applicable, we assume a flat Universe with
$(\Omega_{m},\Omega_{\Lambda},h_0) = (0.3,0.7,0.7)$.  In this
cosmology, an arcsecond corresponds to 7.1\,kpc at $z=4$.

\begin{figure}
 \centerline{\includegraphics[width=3.5in]{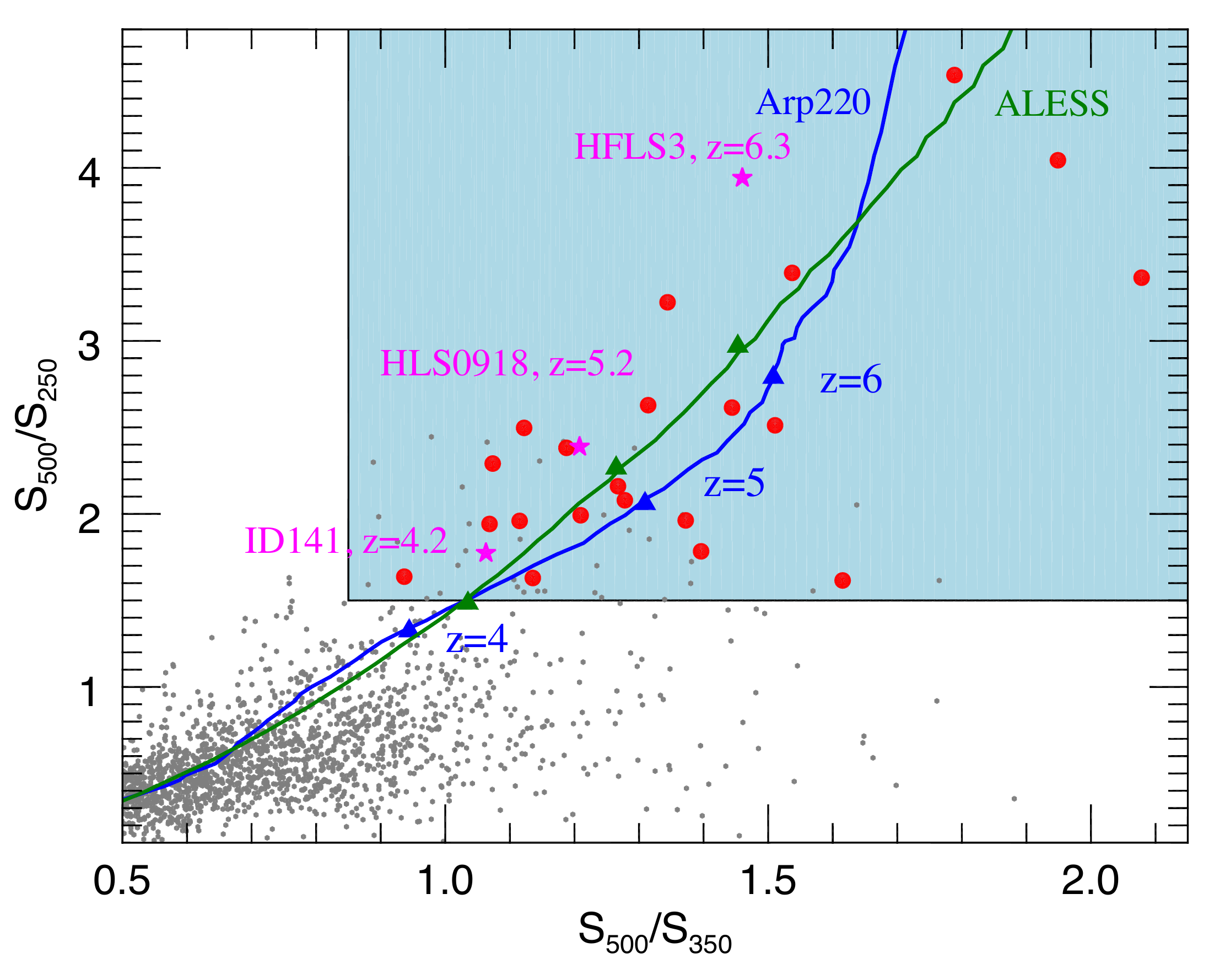}}
 \caption{$S_{500}/S_{350}$ versus $S_{500}/S_{250}$ plot of our 21 
   targets (red circles), and SPIRE-selected DSFGs at $z>4$
   previously studied \citep[magenta stars, 
   ][]{cox11,combes12,riechers13}. A sample of 1,000 randomly 
   selected {\it H}-ATLAS galaxies from \citet{valiante16} are 
   shown as grey dots. We also show the redshift tracks of 
   Arp\,220 (blue line) and of a spectral energy distribution 
   (SED) synthesized from 122 high-redshift DSFGs \citep[green 
   line;][]{cunha15} where triangles indicate $z=4$, 5 and 6. Our 
   targets satisfy the ultrared color cuts, 
   $S_{500}/S_{350} \ge 0.85$ and $S_{500}/S_{250} \ge 1.5$ (blue 
   shaded area), expected for $z\gs 4$ DSFGs \citep{ivison16}.}
\label{ctrack}
\end{figure}

\section{Sample selection}
\label{sample}

\begin{table}
	\begin{center}
	\caption{Targets for which 3-mm spectral scans were obtained.}
	\label{target}
	\begin{tabular}{llc} 
		\hline\hline
		 Nickname & IAU name$^{\rm a}$ \\
		\hline
		 SGP-196076$^{\rm b}$ (SGP-38326$^{\rm c}$)&HATLAS\,J000306.9$-$330248\\
		 SGP-261206$^{\rm b}$&HATLAS\,J000607.6$-$322639\\
		 SGP-354388$^{\rm b,d}$&HATLAS\,J004223.5$-$334340\\
		 SGP-32338$^{\rm b}$&HATLAS\,J010740.7$-$282711\\
		 G09-59393$^{\rm e}$&HATLAS\,J084113.6$-$004114\\
		 G09-81106$^{\rm e}$&HATLAS\,J084937.0+001455\\
		 G09-83808$^{\rm e}$&HATLAS\,J090045.4+004125\\
		 G09-62610$^{\rm e}$&HATLAS\,J090925.0+015542\\
		 G15-26675&HATLAS\,J144433.3+001639\\
		 G15-82684$^{\rm e}$&HATLAS\,J145012.7+014813\\
		 NGP-206987$^{\rm e}$&HATLAS\,J125440.7+264925\\
		 NGP-111912$^{\rm e}$&HATLAS\,J130823.9+254514\\
		 NGP-136156$^{\rm e}$&HATLAS\,J132627.5+335633\\
		 NGP-126191$^{\rm e}$&HATLAS\,J133217.4+343945\\
		 NGP-284357&HATLAS\,J133251.5+332339\\
		 NGP-190387$^{\rm e}$&HATLAS\,J133337.6+241541\\
		 NGP-113609$^{\rm e}$&HATLAS\,J133836.0+273247\\
		 NGP-252305$^{\rm e}$&HATLAS\,J133919.3+245056\\
		 NGP-63663$^{\rm e}$&HATLAS\,J134040.3+323709\\
		 NGP-246114$^{\rm e}$&HATLAS\,J134114.2+335934\\
		 NGP-101333$^{\rm e}$&HATLAS\,J134119.4+341346\\
		\hline
	\end{tabular}
	\end{center}
	$\rm^{a}$ As listed in \citet{ivison16}.\\
	$\rm^{b}$ Observed with ALMA at 3\,mm.\\
	$\rm^{c}$ Old nomenclature used by \citet{oteo16}.\\
	$\rm^{d}$ Also known as the Great Red Hope \citep{oteo17grh}.\\
	$\rm^{e}$ Observed with NOEMA at 1.3\,mm as well as at 3\,mm (\S\ref{noema}).
\end{table}

 Our targets -- see Table~\ref{target} -- were chosen from the faint,
`ultrared' galaxy sample of \citet{ivison16}, taking those best suited
to the latitudes of the telescopes we employ, with photometric
redshifts consistent with $z\gs4$. Here, we briefly summarize the
selection method used, referring readers to \citet{ivison16} for more
details.

The sample was selected from the SPIRE images used to construct {\it
  H}-ATLAS Data Release 1 \citep{valiante16}, employing an optimal
extraction kernel to minimize the effects of source confusion, which
is especially pernicious at 500\,$\mu$m. The reddest galaxies were
isolated based on their SPIRE colors, such that
$S_{500}/S_{250}\ge1.5$ and $S_{500}/S_{350}\ge0.85$, where $S_{250}$
is the flux density measured at 250\,$\mu$m (see Fig.~\ref{ctrack}).
The galaxies thus selected have a median $S_{500}\sim 50$\,mJy, such
that the majority are not expected to be lensed gravitationally
\citep[e.g.][but see \citeauthor{oteo17morph}]{negrello10,conley11}.

The reddest of these SPIRE-selected galaxies were then imaged
with SCUBA-2 \citep{holland13} on the 15-m JCMT and/or with
LABOCA \citep{siringo09} on the 12-m Atacama Pathfinder Telescope
(APEX) so that better photometric redshifts could be determined.
These data are also utilized here, in \S\ref{continuum}, to aid
us in spatially localizing any line emission.  Of the 109 objects
thus targeted by \citeauthor{ivison16}, 17 galaxies were selected
for further observations with the Institute Radioastronomie
Millimetrique's (IRAM's) Northern Extended Millimeter Array
(NOEMA) and four galaxies for further observations with the
Atacama Large Millimeter Array (ALMA), based on their
accessibility to those telescopes and their high photometric
redshifts. The SPIRE flux densities and photometric redshifts
determined by \citet{ivison16} are listed in Table~\ref{mescont}.

\section{Observations}
\label{obs}

\begin{figure*}
 \includegraphics[width=0.185\textwidth]{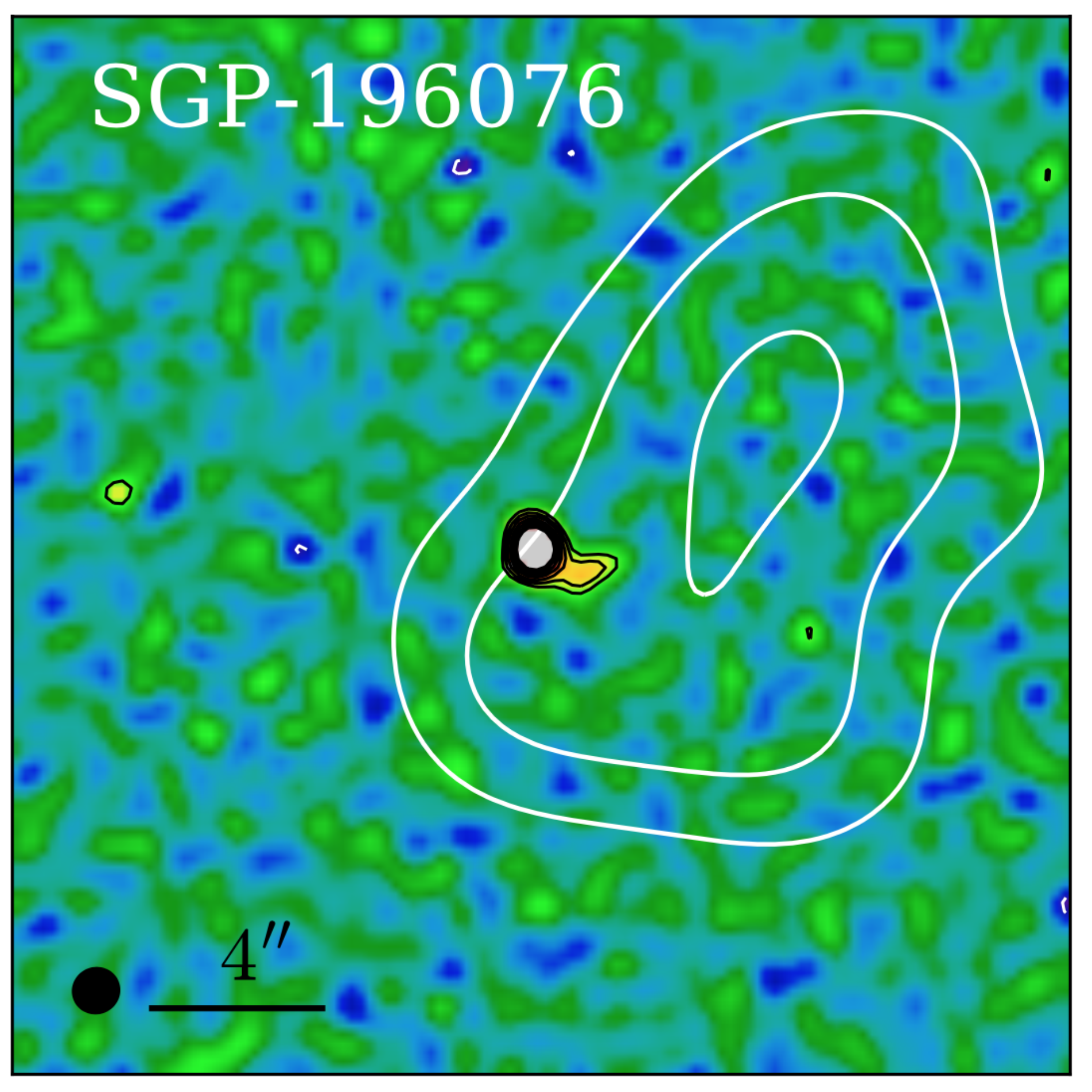}
 \hspace*{6mm}
 \includegraphics[width=0.185\textwidth]{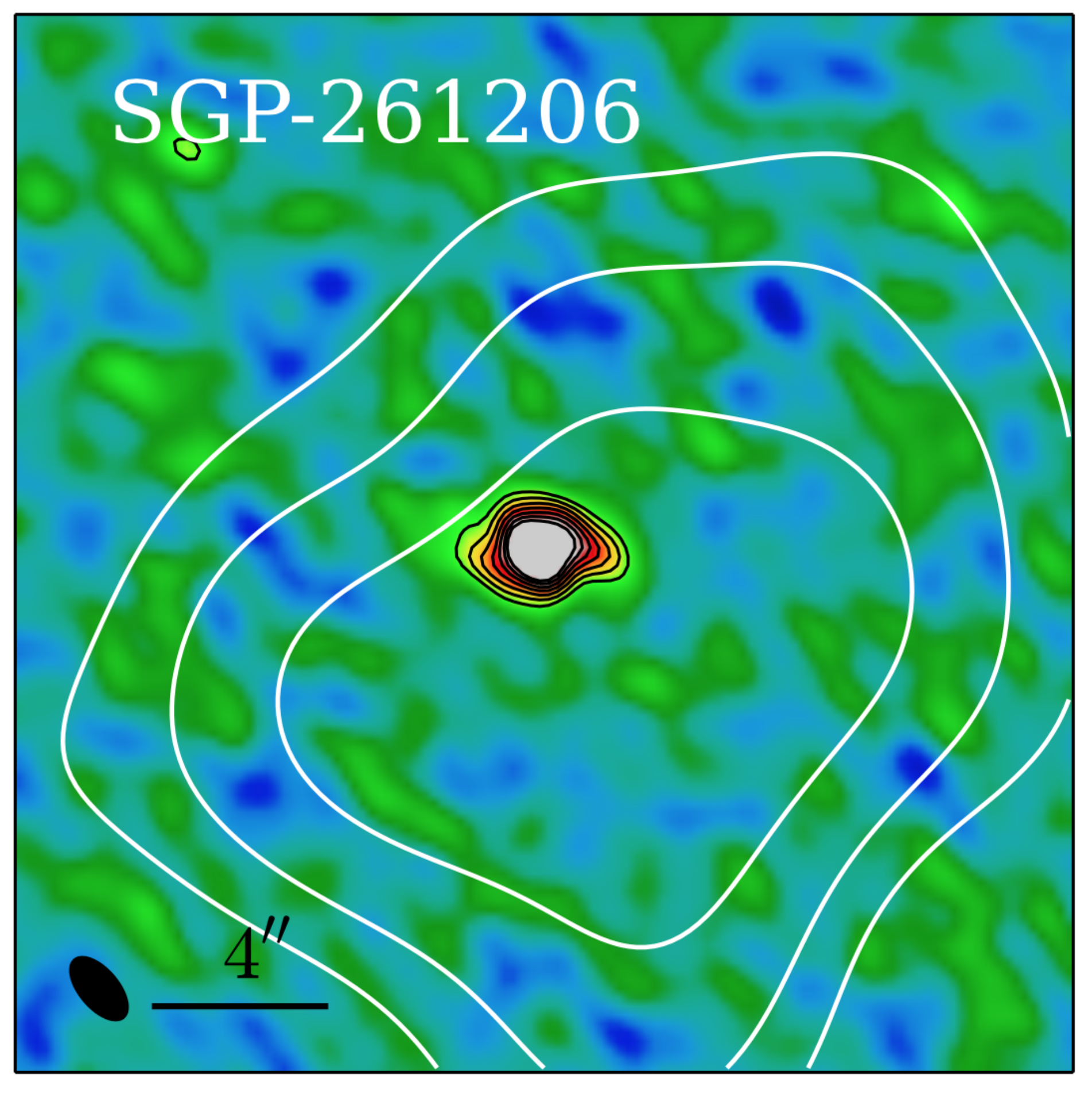}
\hspace*{6mm}
 \includegraphics[width=0.185\textwidth]{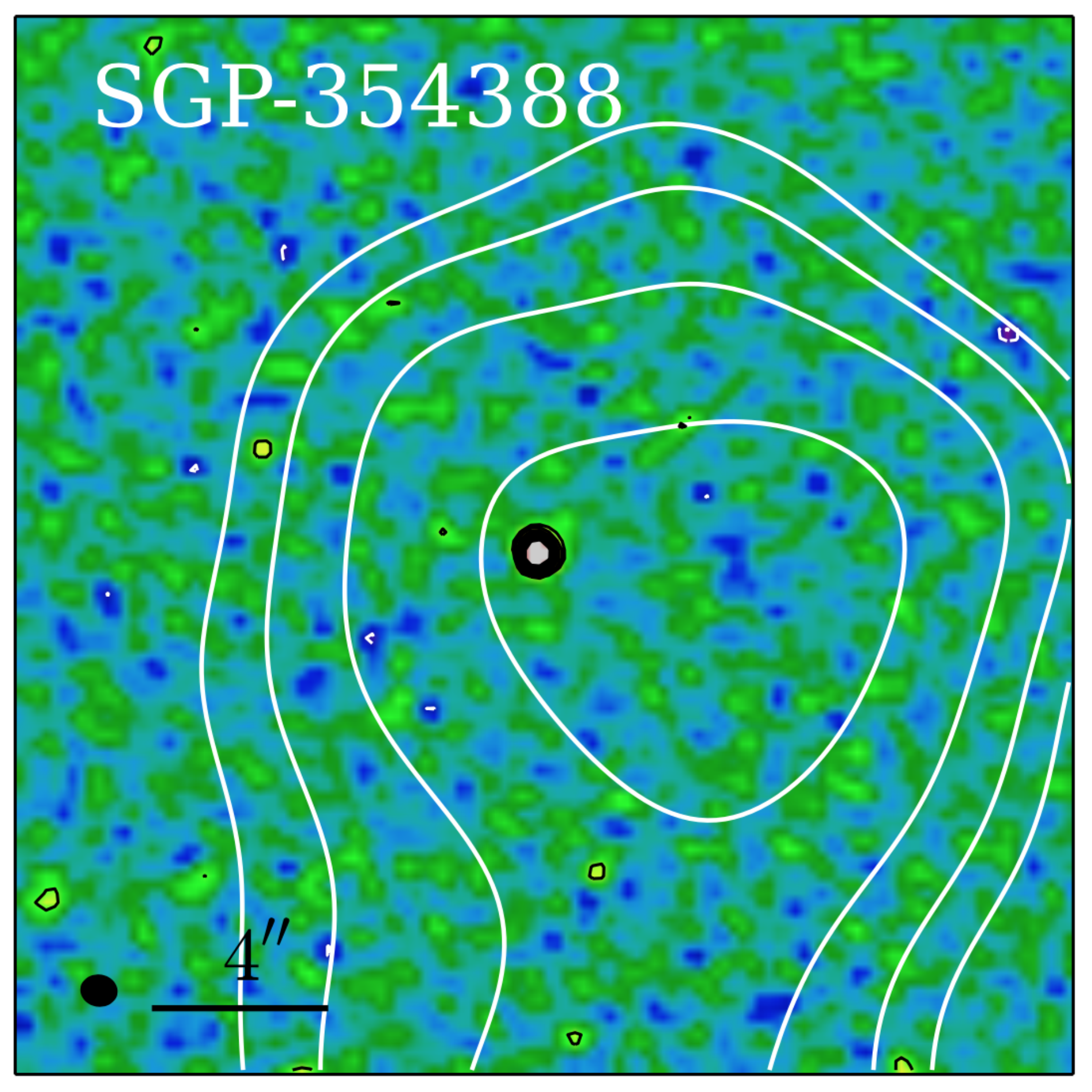}
\hspace*{6mm}
 \includegraphics[width=0.185\textwidth]{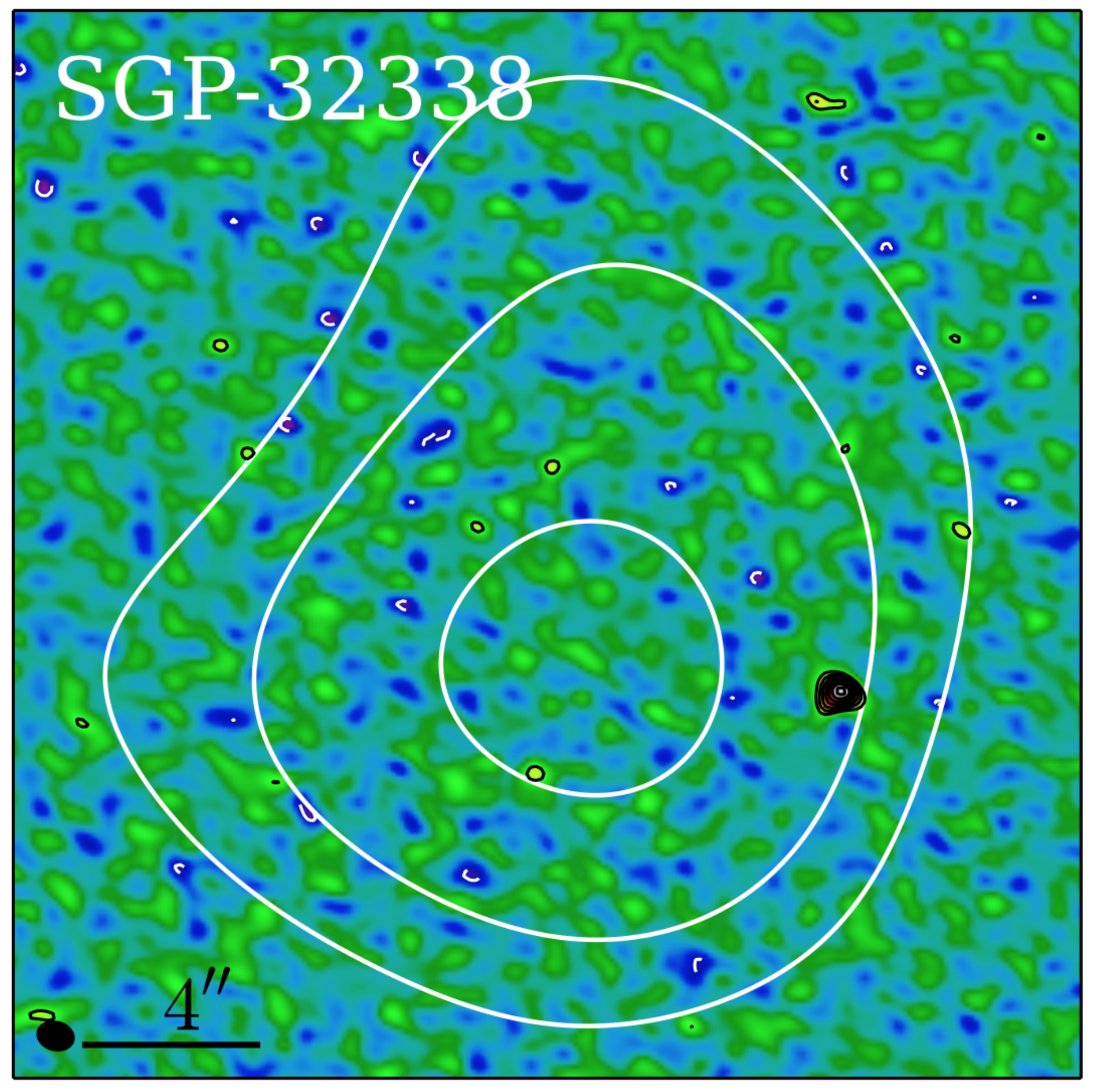}
 
 \includegraphics[width=0.225\textwidth]{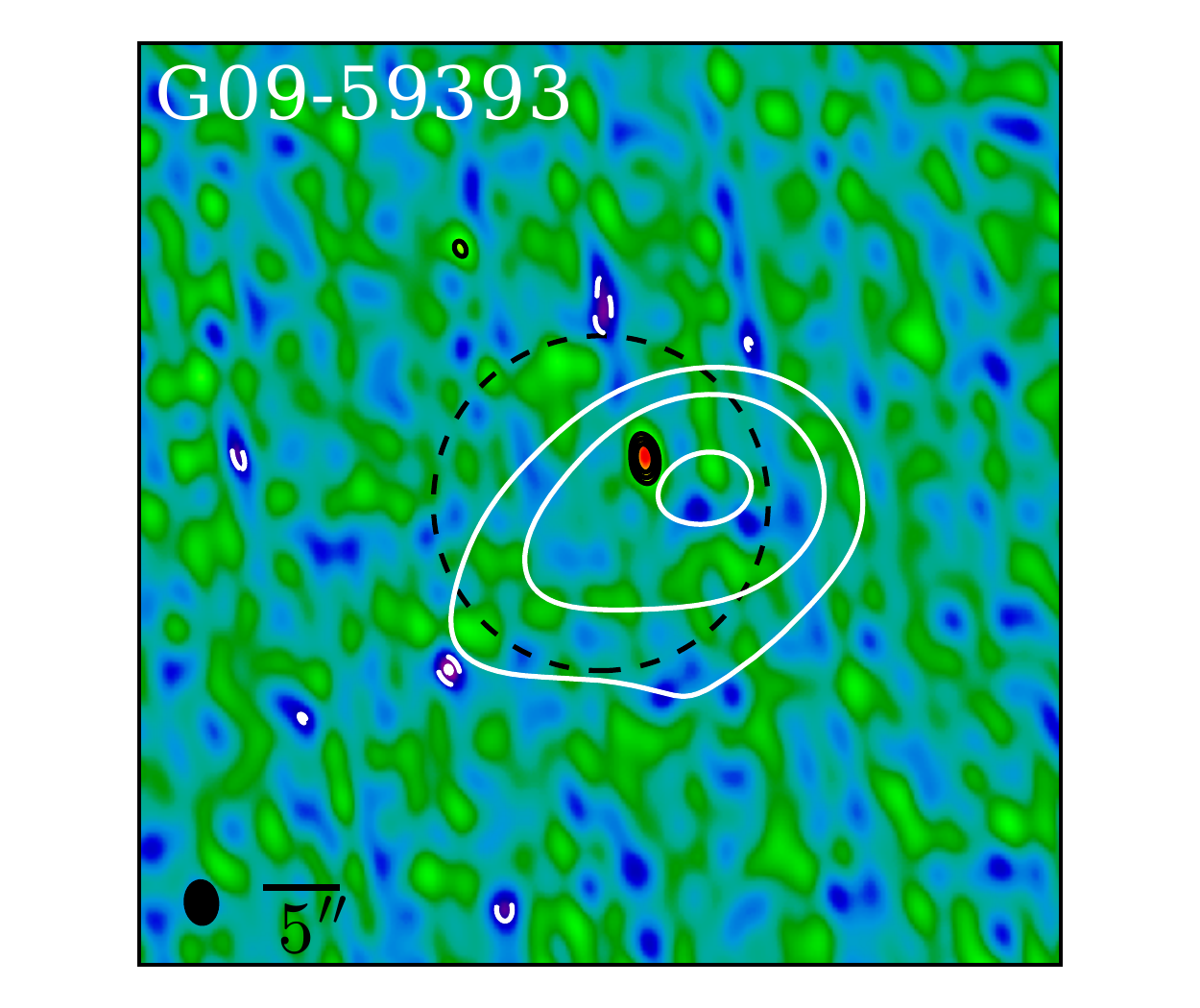}
 \includegraphics[width=0.225\textwidth]{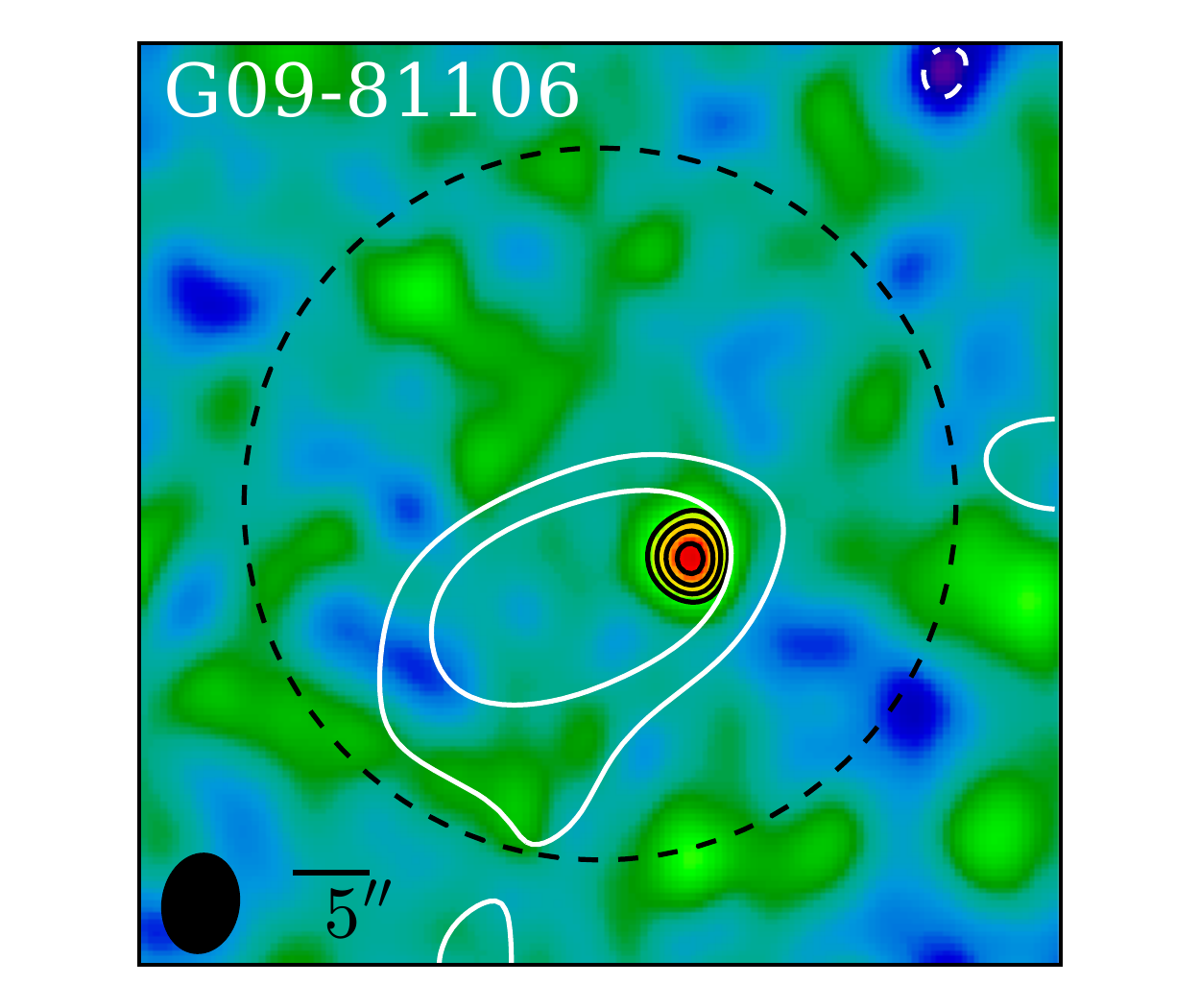}
 \includegraphics[width=0.225\textwidth]{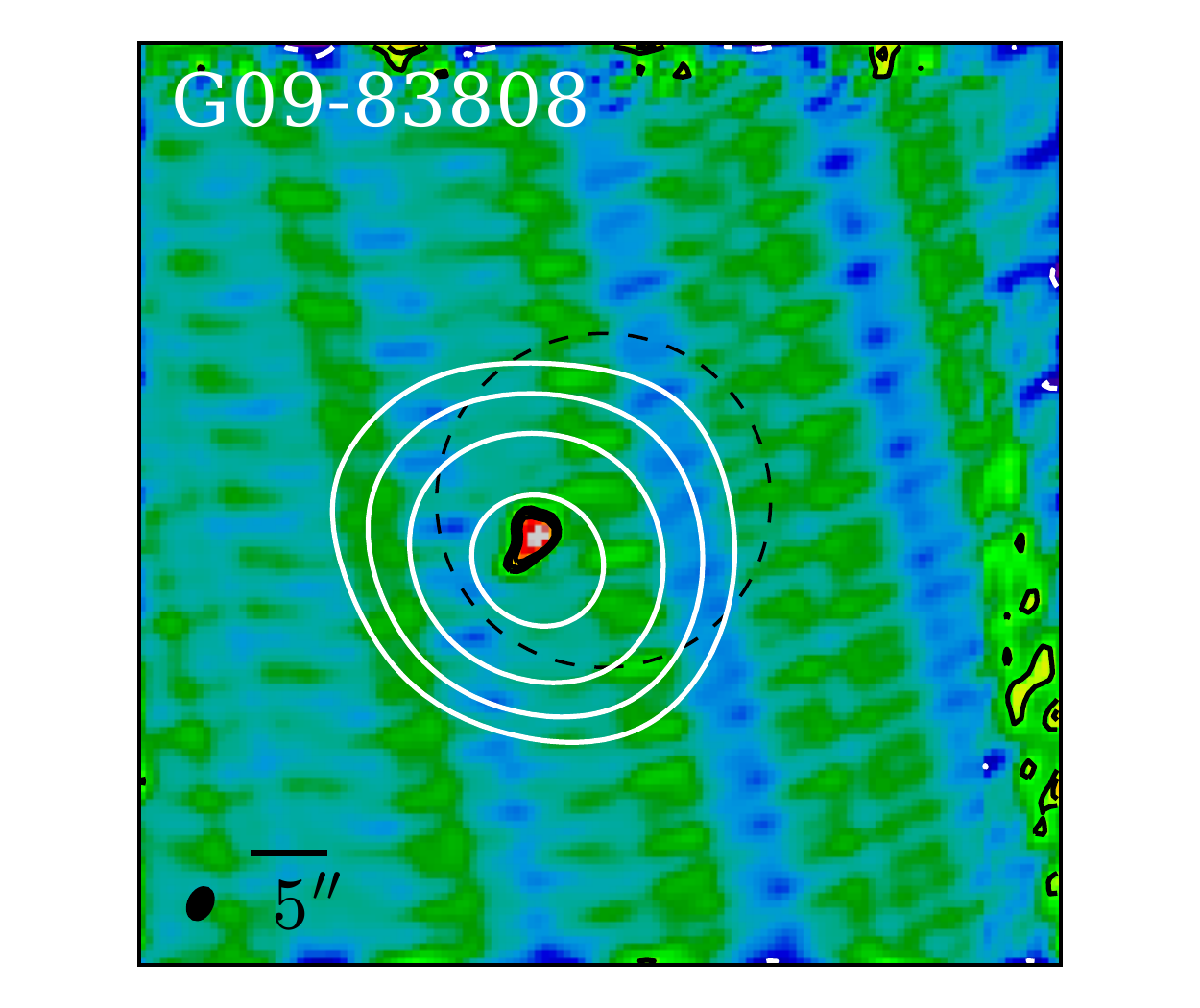}
 \includegraphics[width=0.225\textwidth]{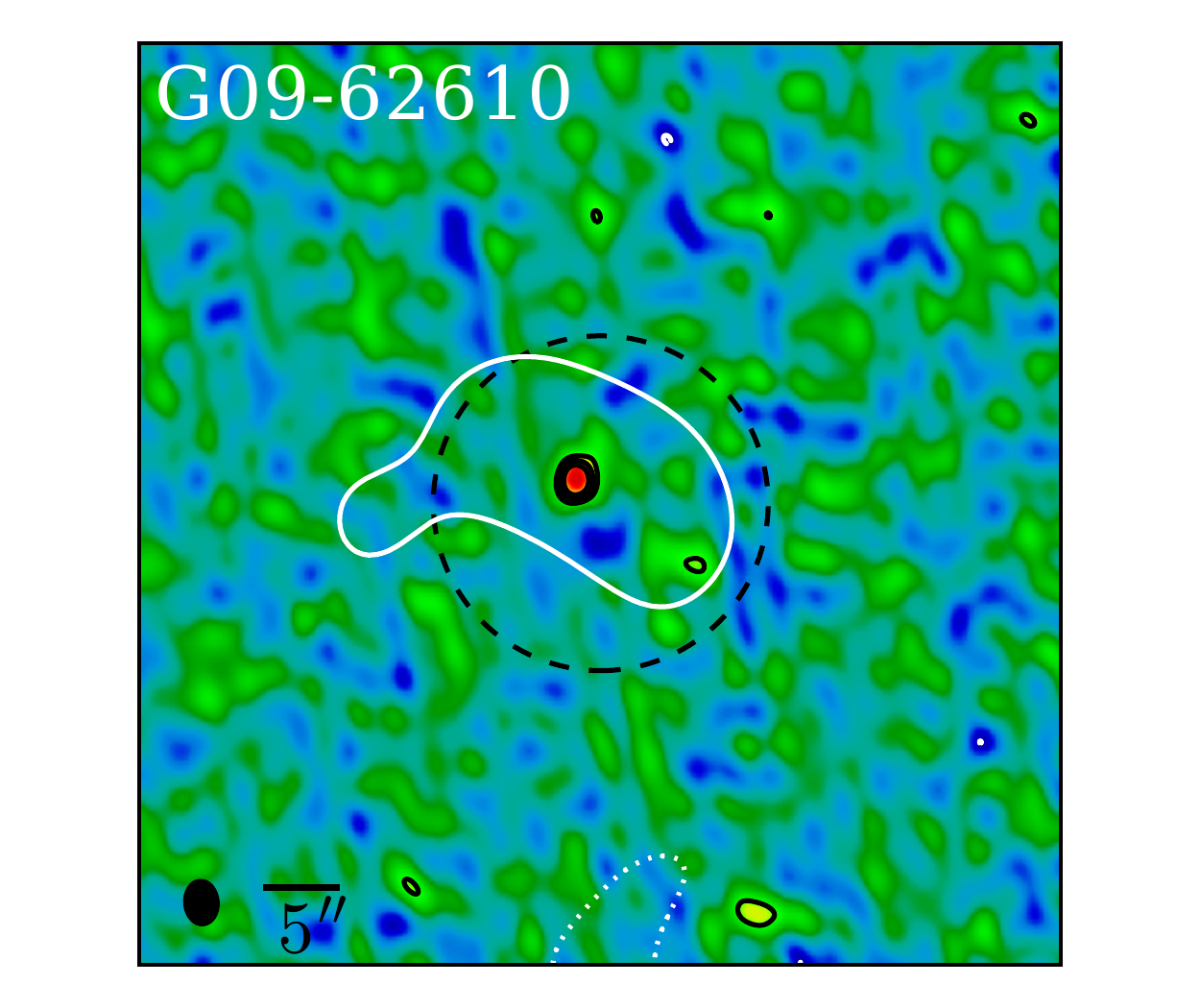}

 \includegraphics[width=0.225\textwidth]{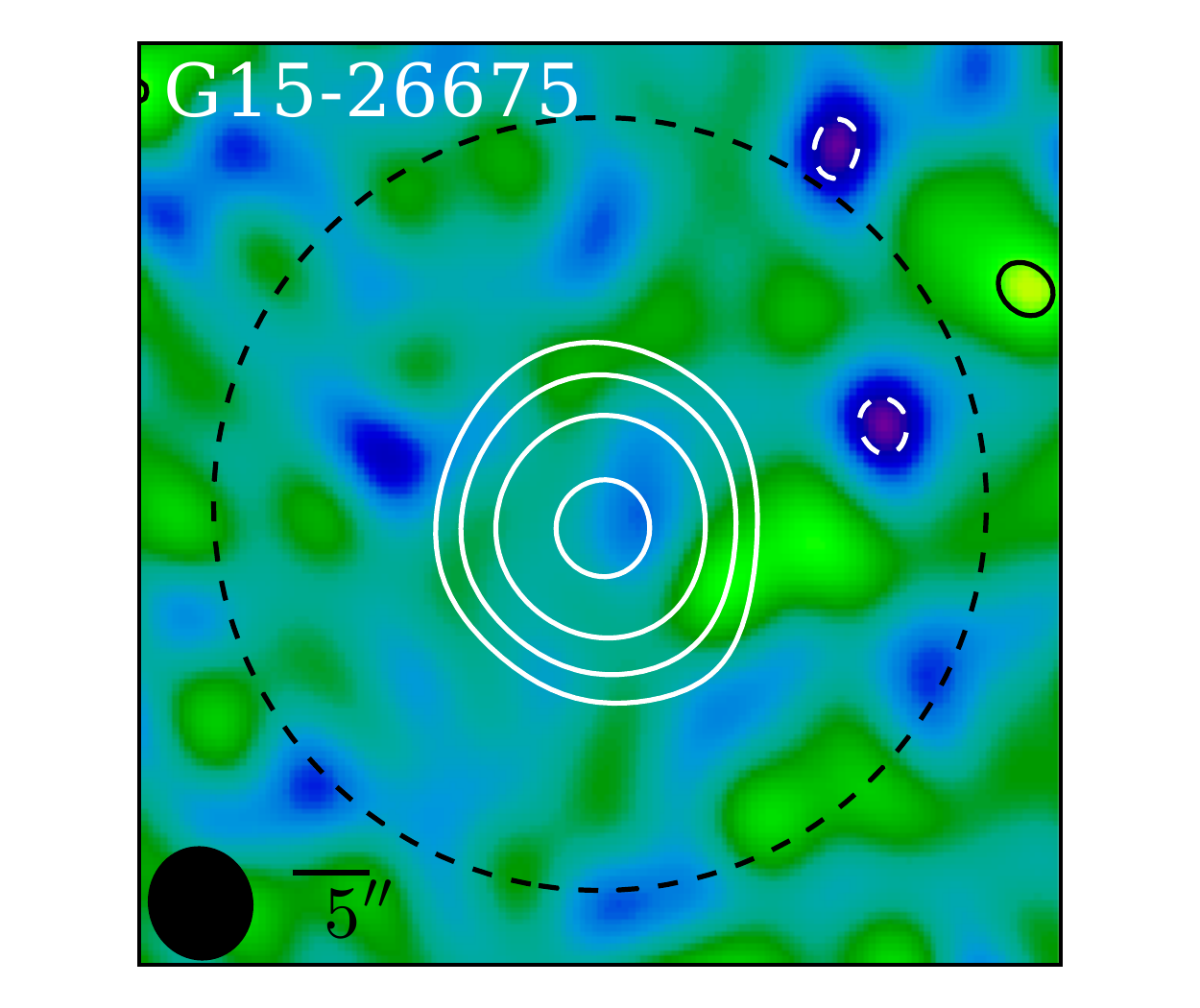}
 \includegraphics[width=0.225\textwidth]{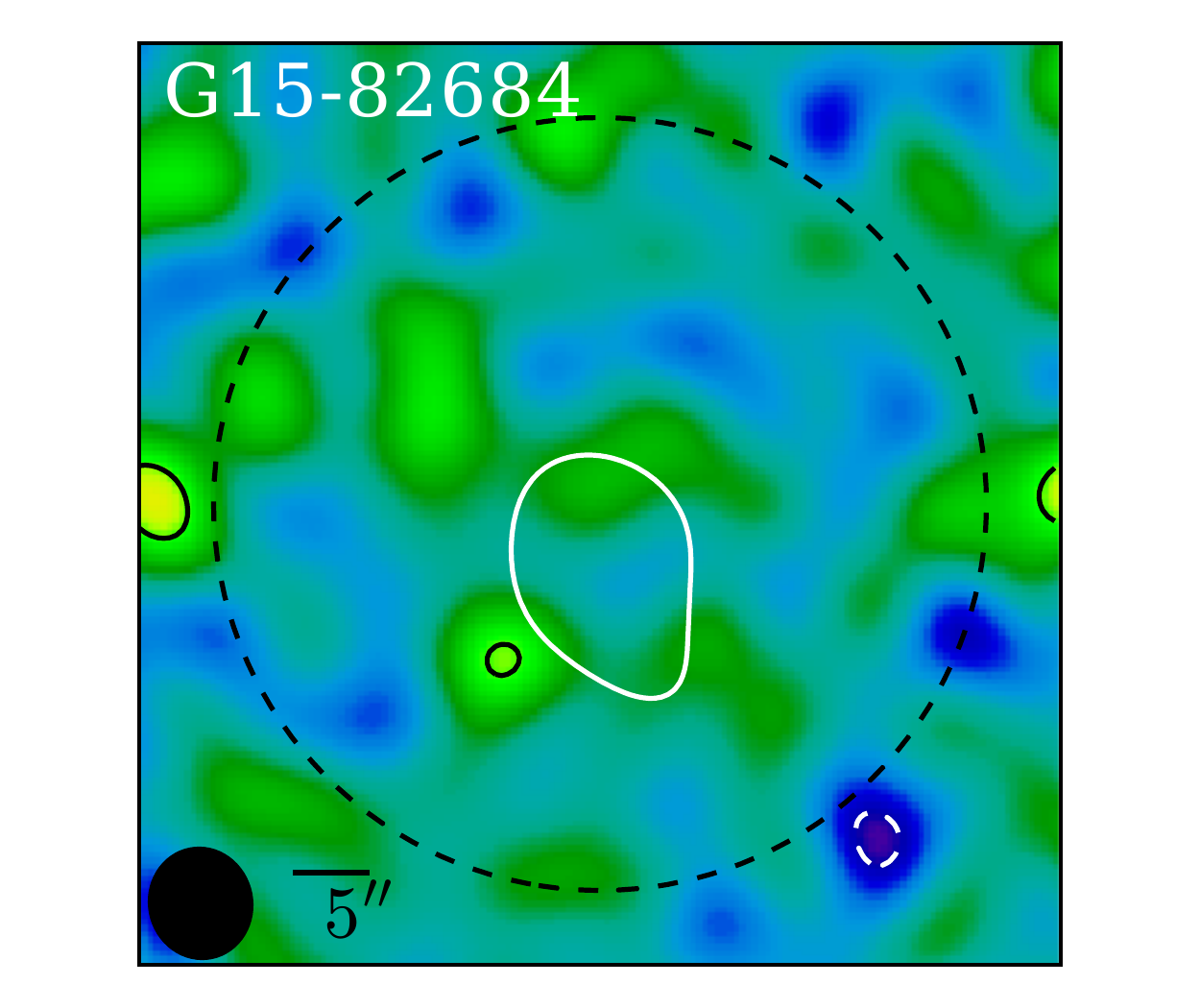}
 \includegraphics[width=0.225\textwidth]{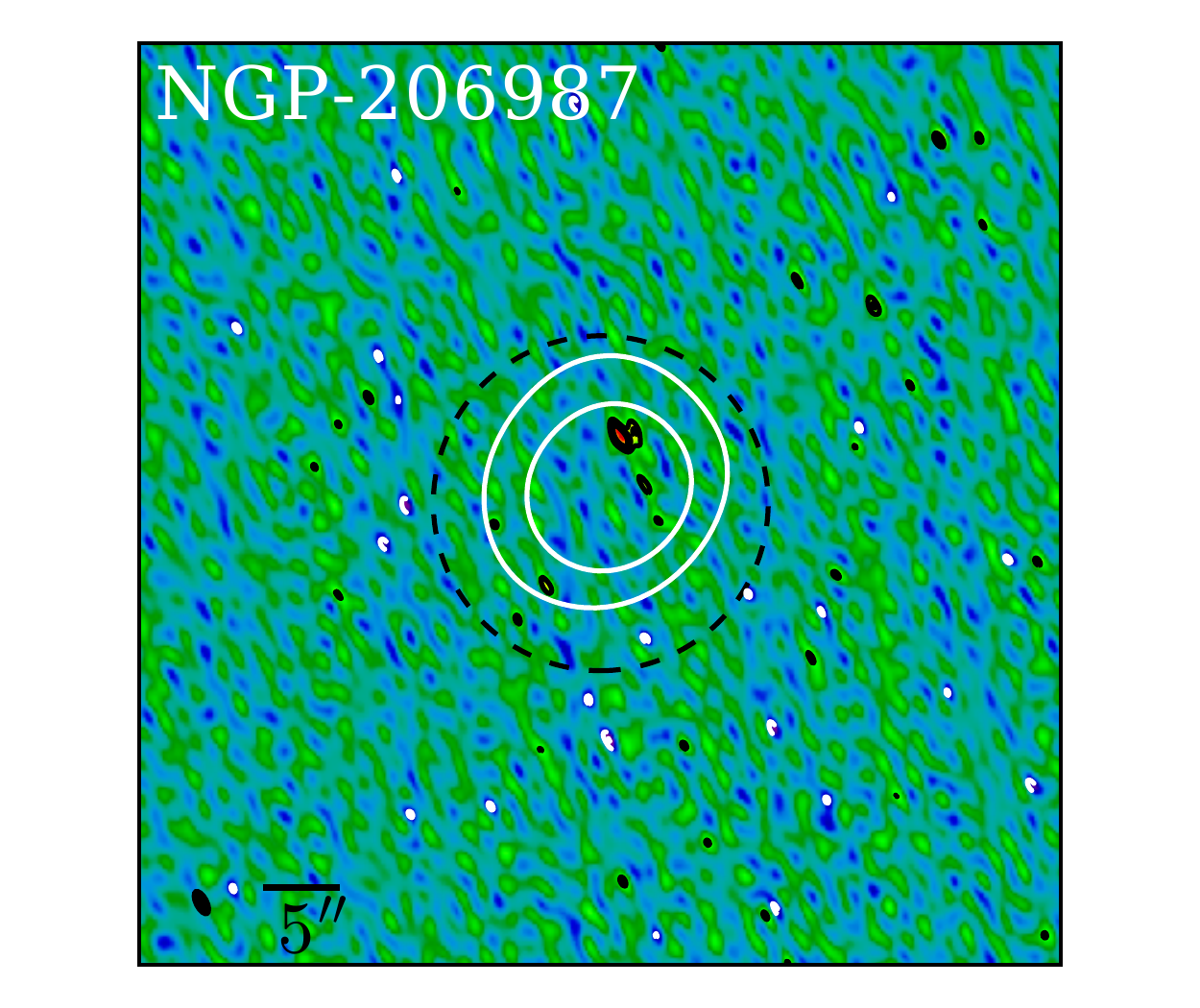}
 \includegraphics[width=0.225\textwidth]{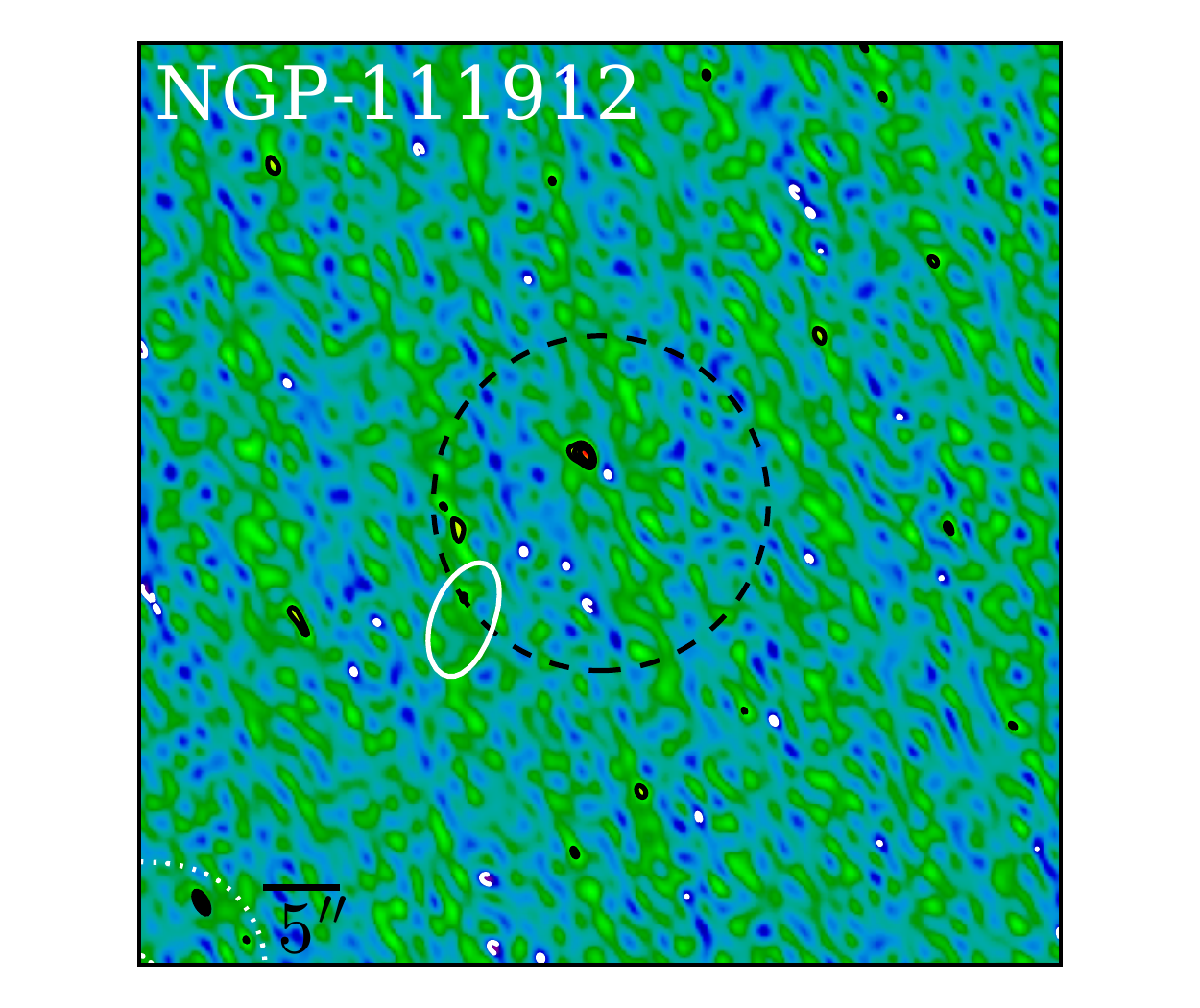}

 \includegraphics[width=0.225\textwidth]{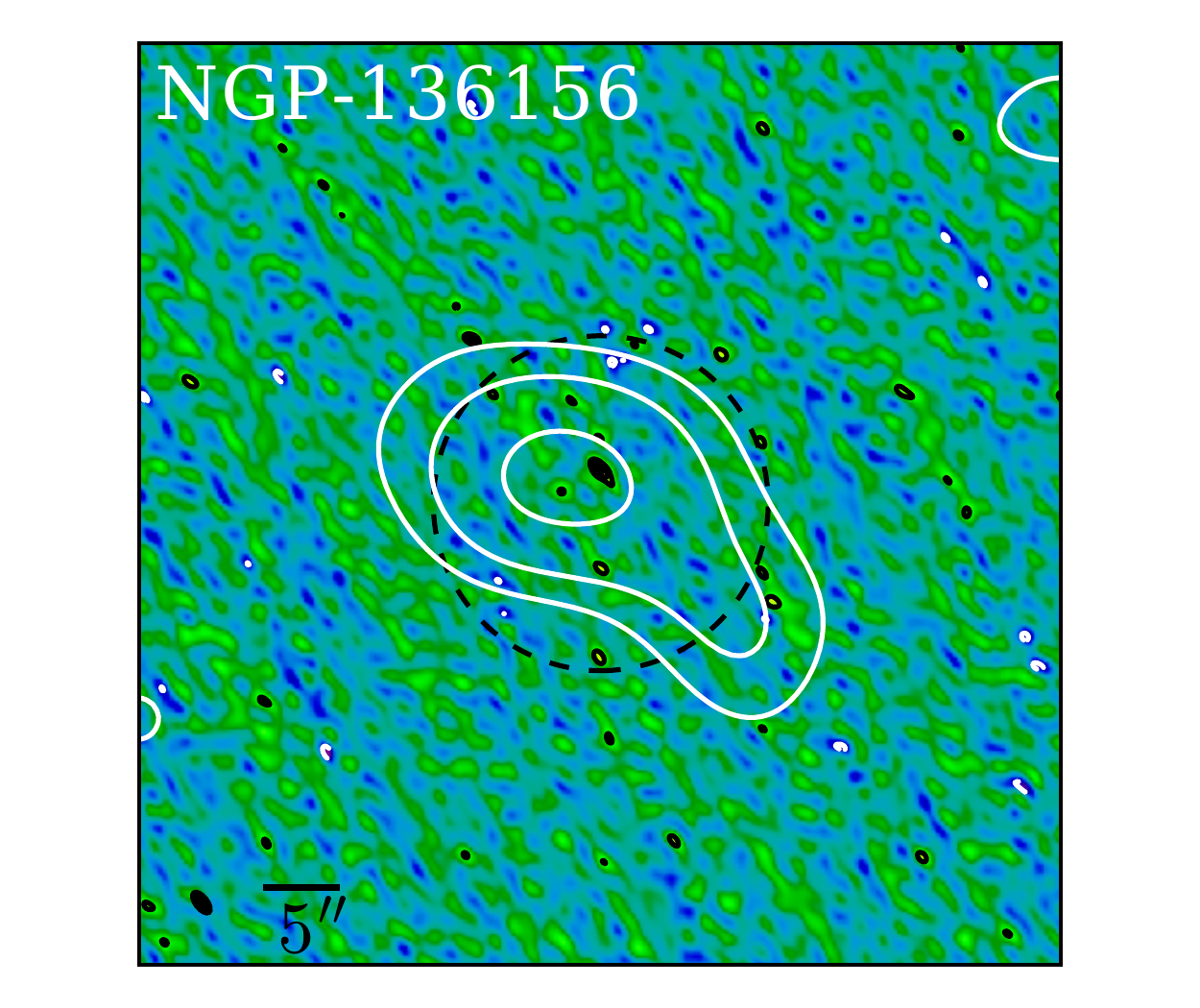}
 \includegraphics[width=0.225\textwidth]{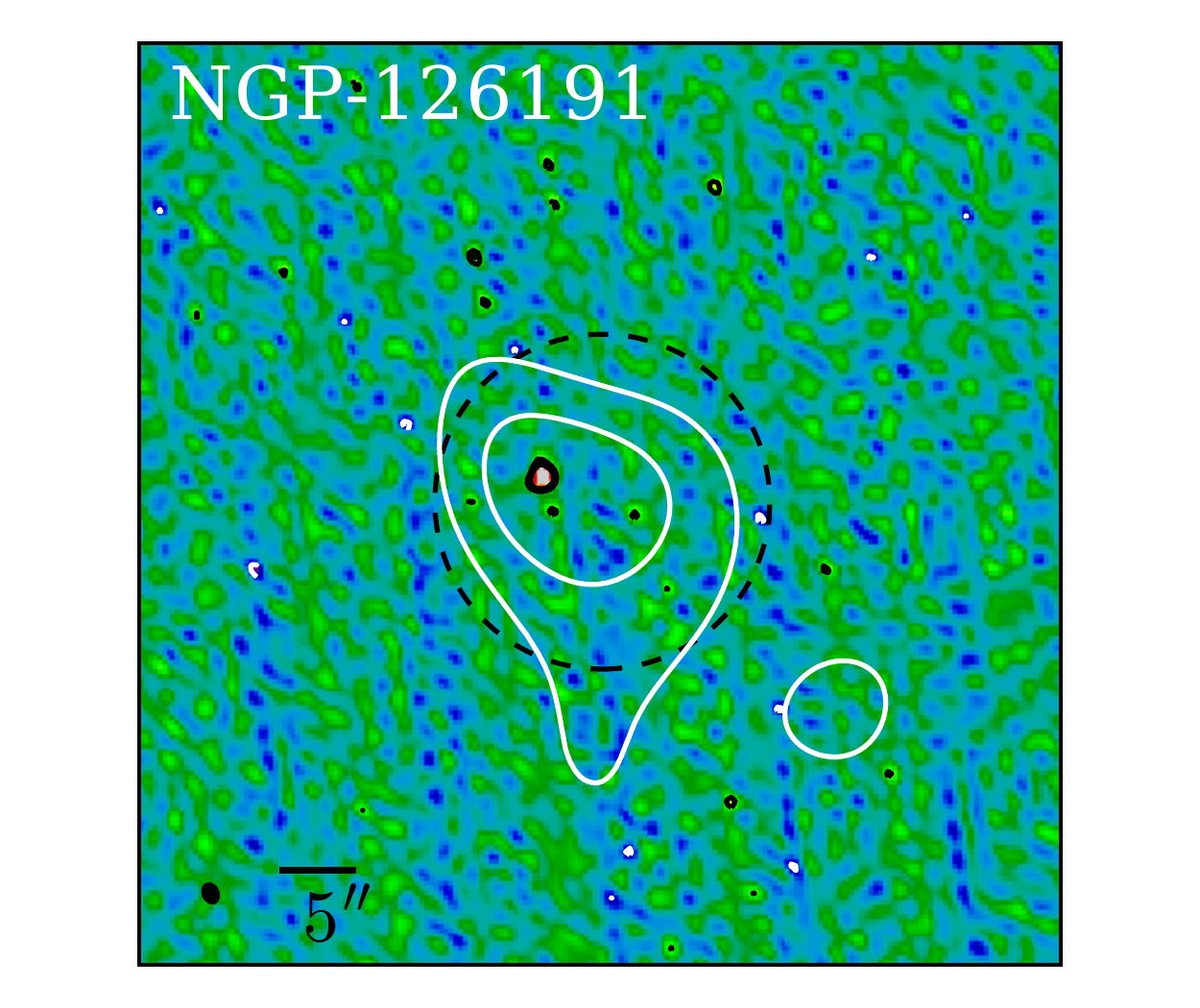}
 \includegraphics[width=0.2255\textwidth]{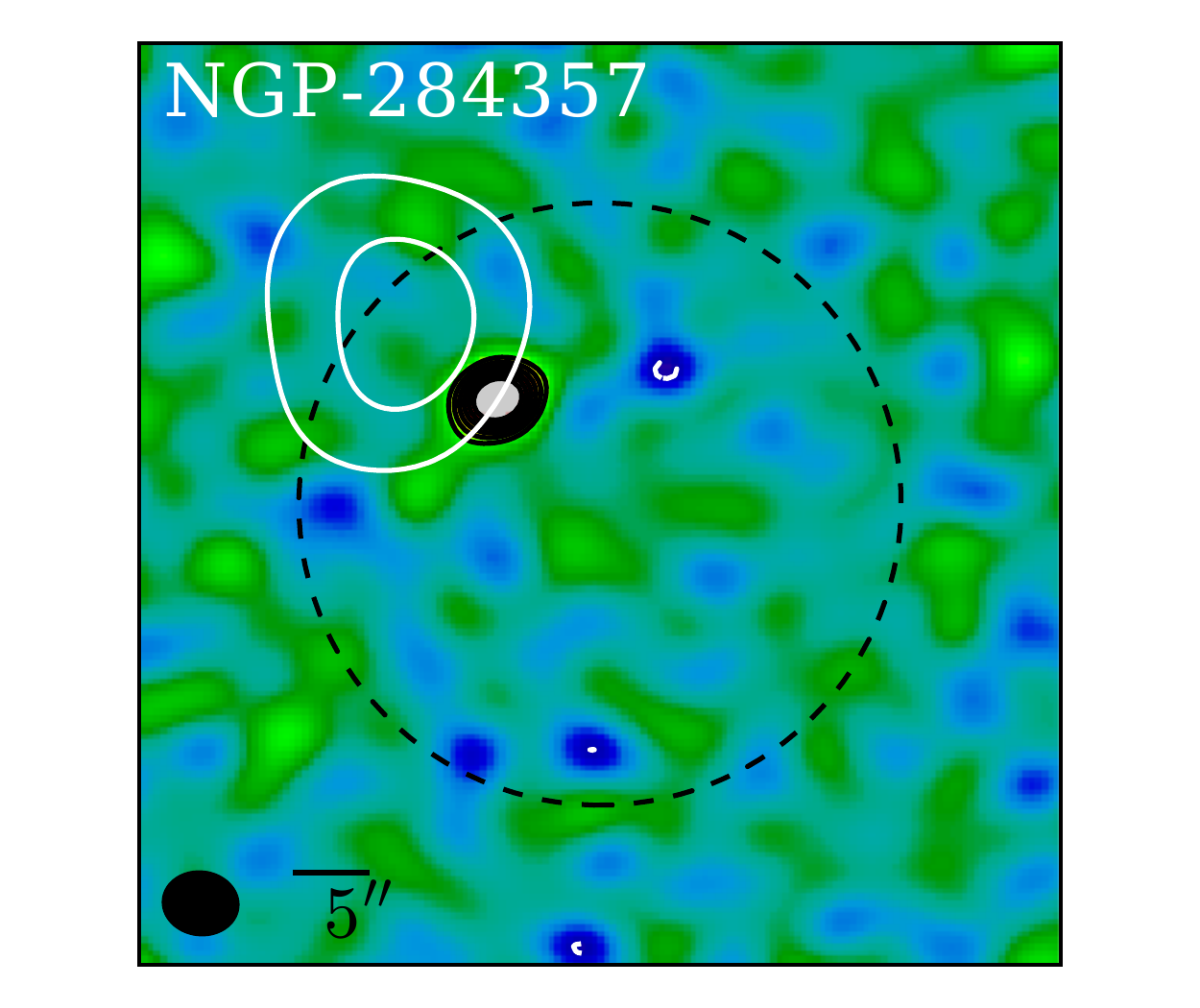}
 \includegraphics[width=0.225\textwidth]{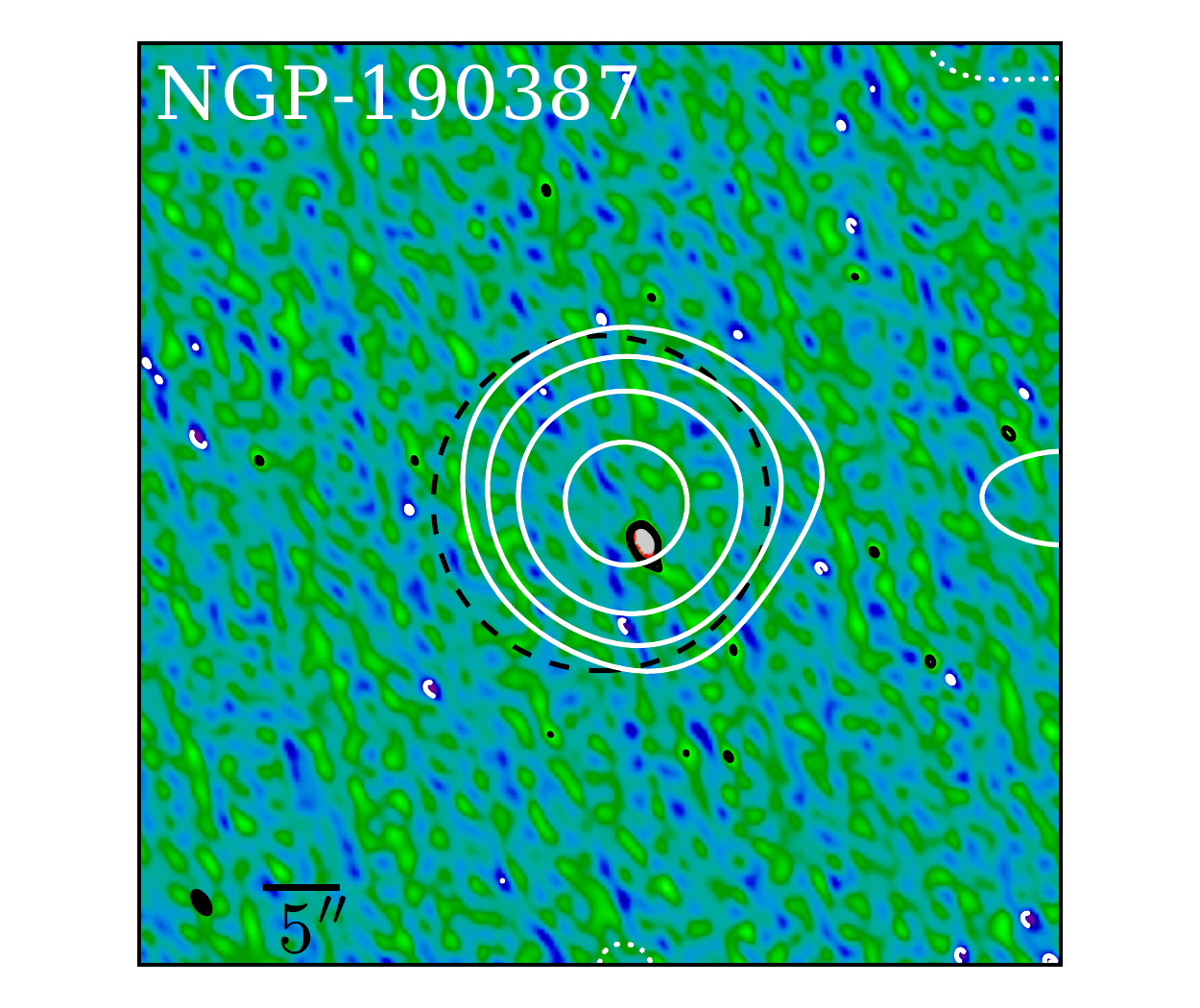}

 \includegraphics[width=0.225\textwidth]{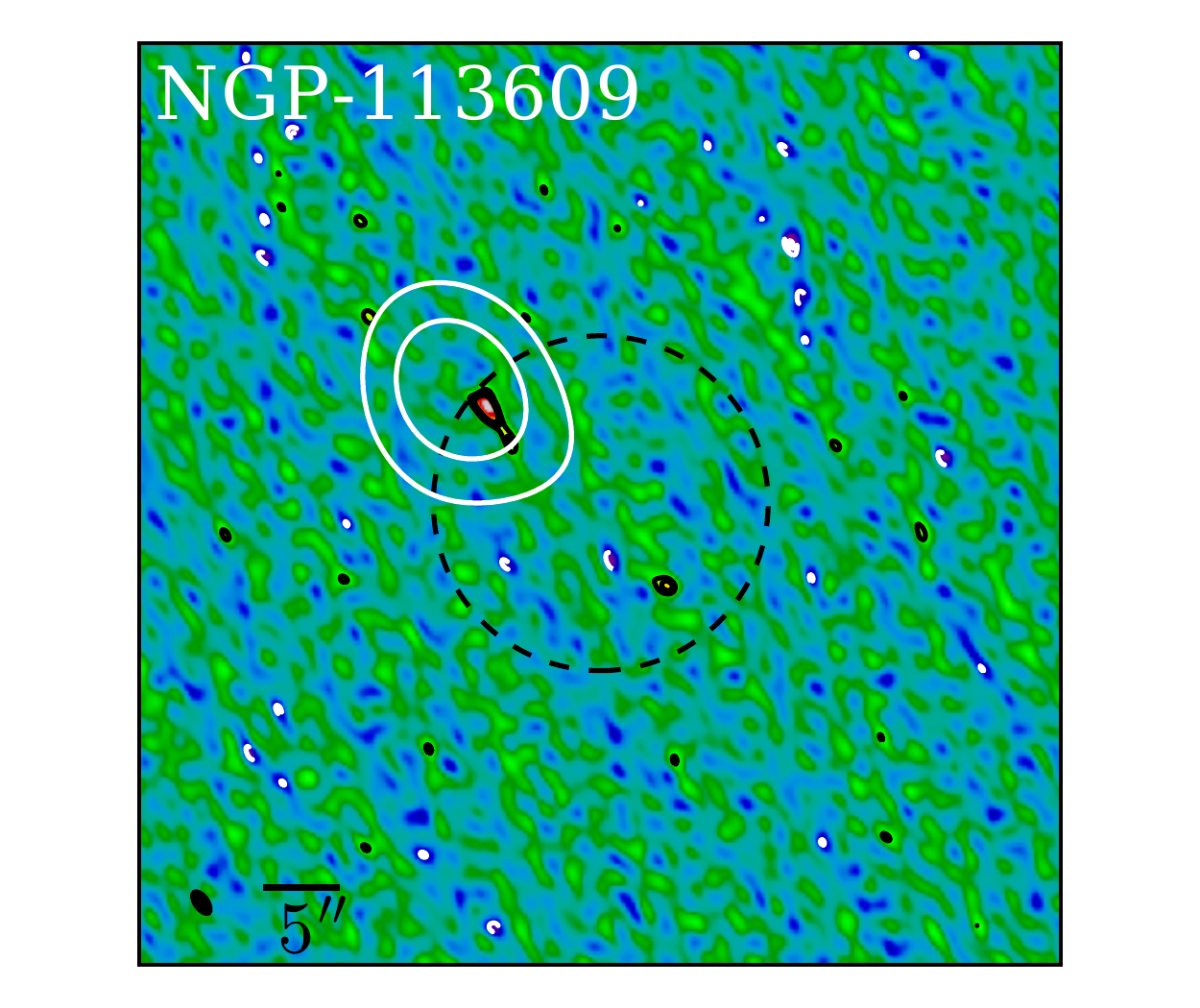}
 \includegraphics[width=0.225\textwidth]{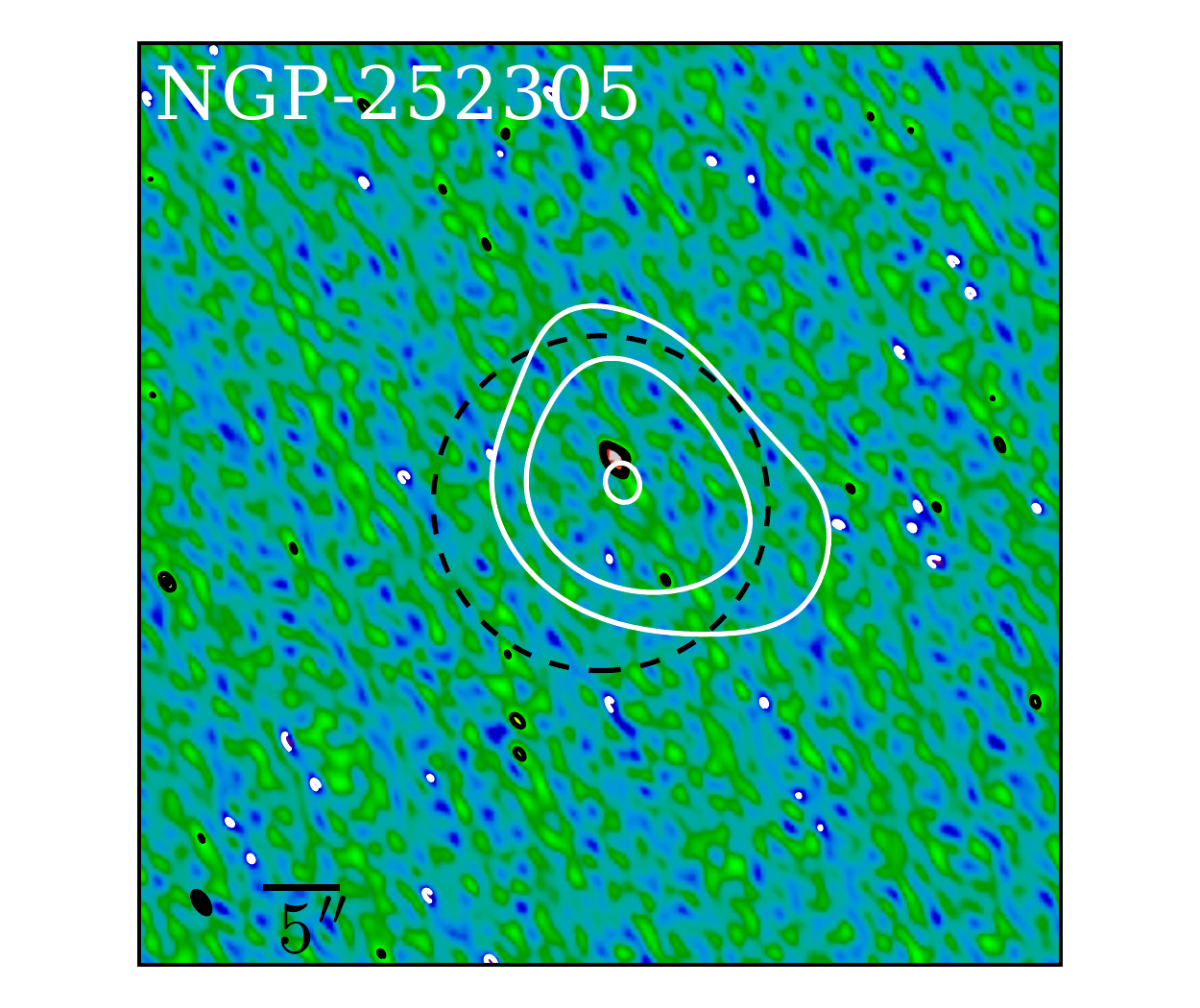}
 \includegraphics[width=0.225\textwidth]{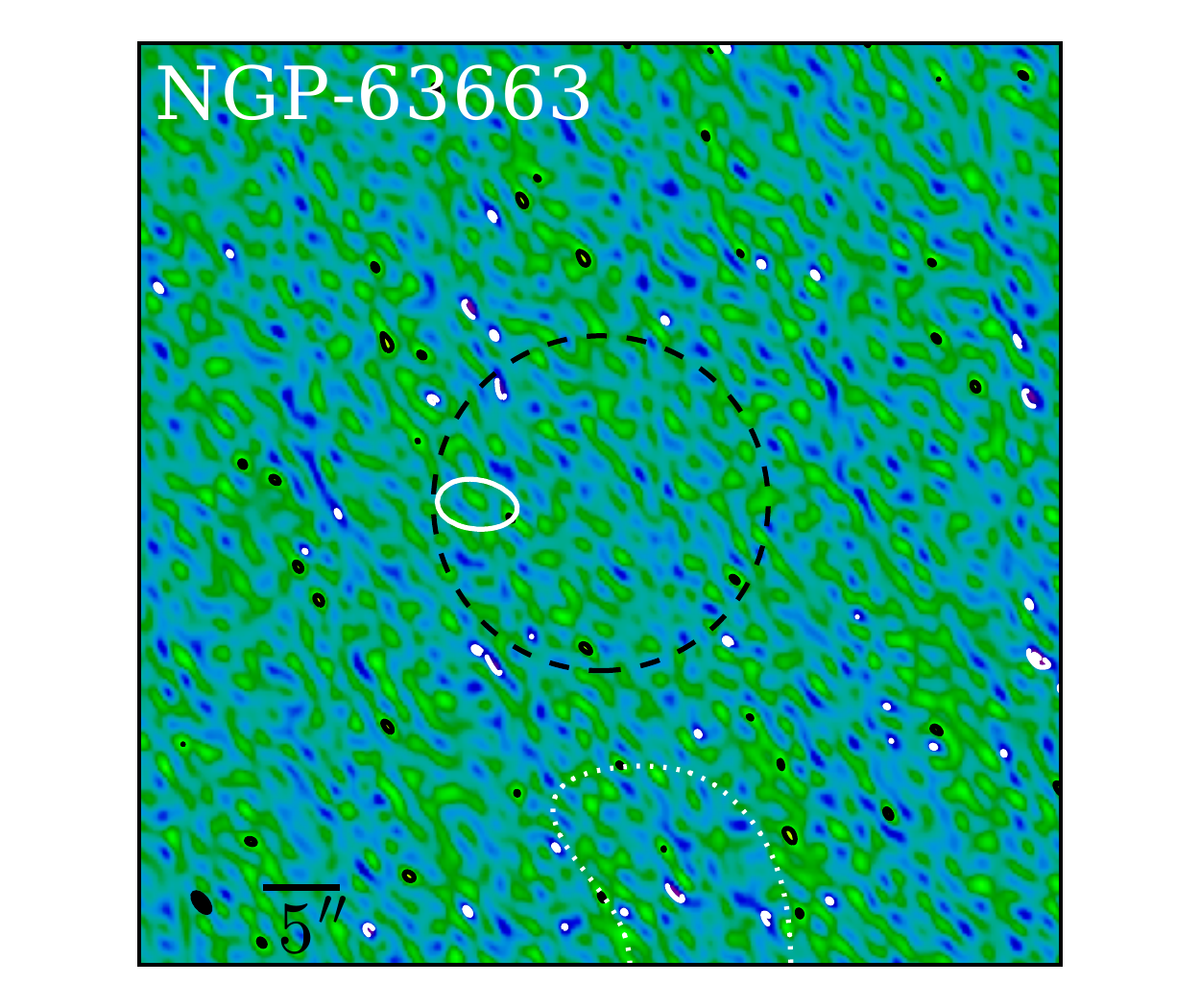}
 \includegraphics[width=0.225\textwidth]{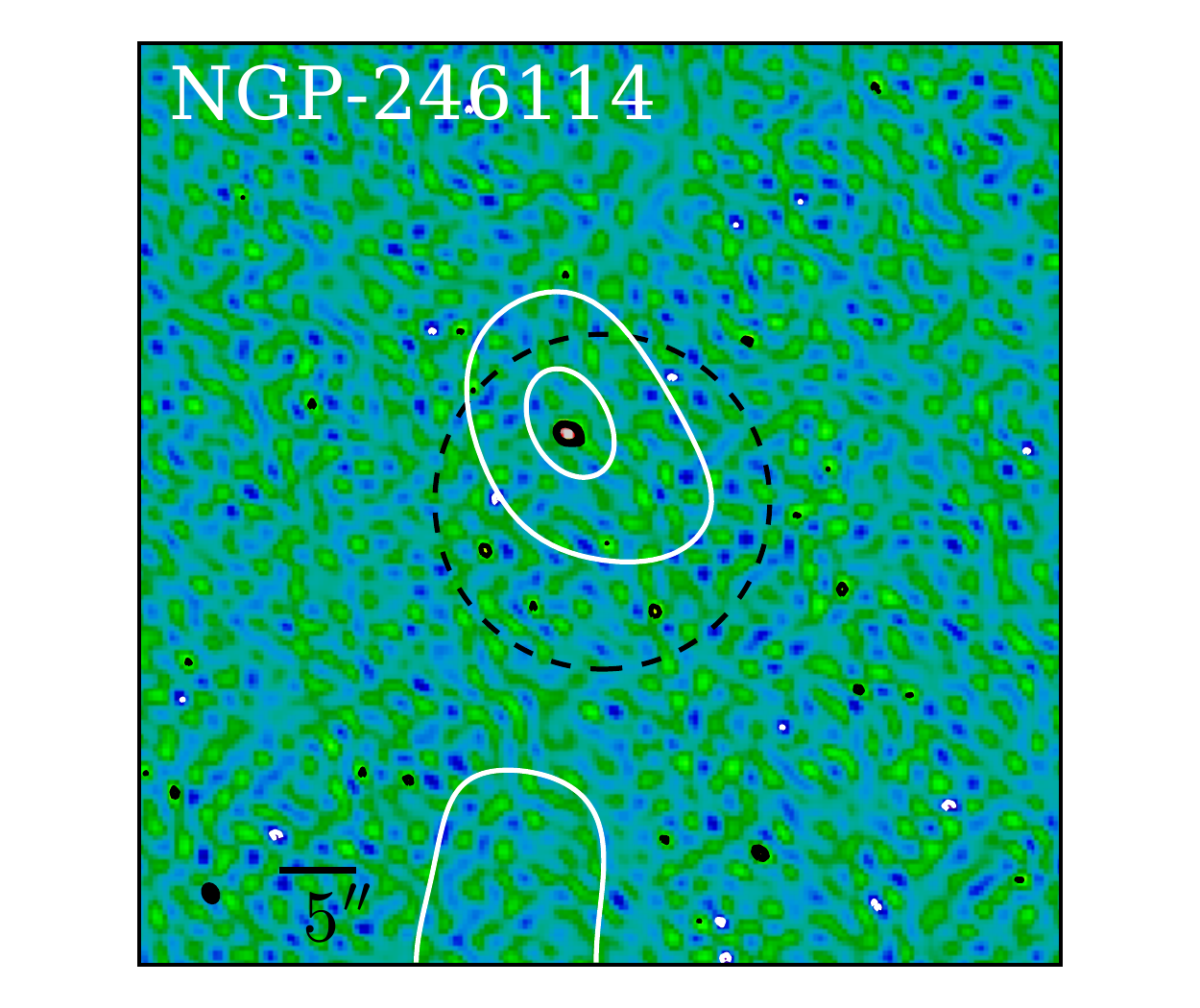}

 \includegraphics[width=0.24\textwidth]{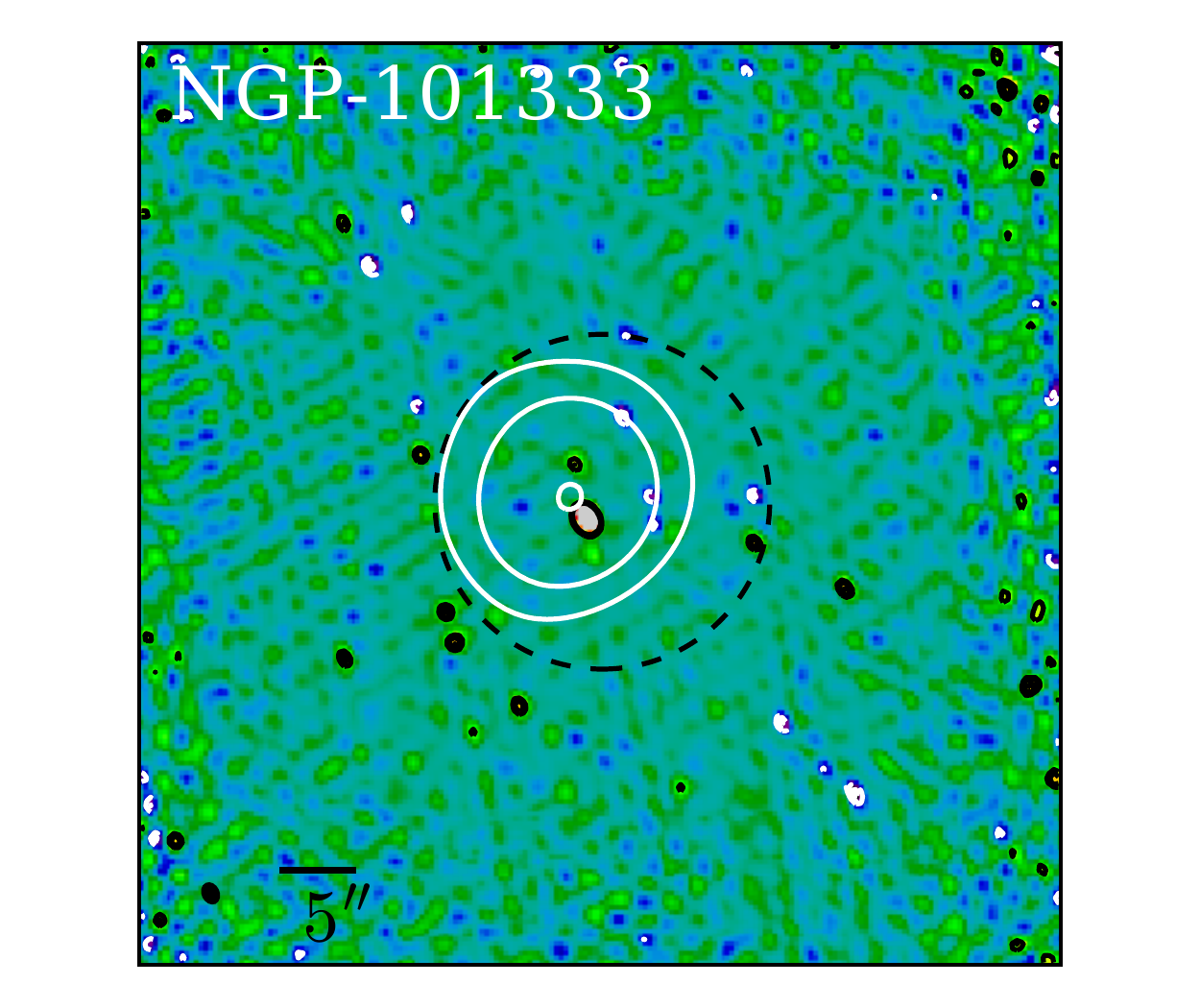}
 \caption{Continuum images of the 1.3- or 3-mm continuum data
   from ALMA {\it (top row)} and NOEMA, uncorrected for the
   primary beam response.  Black contours represent the continuum
   emission, and start from $3\sigma$. Black dashed lines
   indicate where the sensitivity (primary-beam response) drops
   to 50\% of the peak response (for ALMA at 3\,mm, this region
   exceeds the size of the maps shown here).  White contours
   represent 850-$\mu$m emission as detected by SCUBA-2, smoothed
   with a 13$''$ Gaussian, with contours starting at $3\sigma$
   and increasing in factors of $\sqrt 2$, from \citet{ivison16}.
   The SCUBA-2 images give an indication of where to expect 1.3-
   or 3-mm continuum emission.  For the ten examples where 1.3-
   or 3-mm continuum is detected, we see typical offsets of
   $\approx 2$--4$''$ between the emission peaks detected by
   SCUBA-2 and the more precisely pinpointed 1.3- or 3-mm peaks
   detected by ALMA or NOEMA, consistent with the
   $\sigma=2$--3$''$ pointing accuracy of the JCMT for a single
   visit to a target, as was usually the case for the SCUBA-2
   images shown here.  Except for NGP-111912, all continuum
   detections are coincident with emission-line detections.
   White dashed contours indicate $-3\sigma$ at 3\,mm. Black
   ellipses indicate the synthesized beamsize. N is up; E is
   left.}
\label{detcont1}
\end{figure*}

\subsection{NOEMA 3-mm spectral scans}

Our observations with
NOEMA\footnote{http://iram-institute.org/EN/noema-project.php} were
conducted as two programs (Program IDs: W05A, X0C6; Co-PIs:
R.\,J.~Ivison, M.~Krips). Table~\ref{target} lists the galaxies
observed.  Both projects acquired data using five or six antennas in
NOEMA's most compact (D) configuration. W05A was carried out between
2012 June and 2013 April, and 14 targets were observed. X0C6 took
place between 2013 November and 2014 June, where four targets were
observed. One target, G09-83808, was observed during both periods.

We employed multiple receiver tunings together with the WideX
correlator -- which provides 3.6\,GHz of instantaneous
dual-polarization bandwidth -- to cover the 80--101.6-GHz part of
the 3-mm atmospheric window, in which we expect to find at least
one $^{12}$CO transition for galaxies at $z>3.6$ -- see, for
example, Figure~2 of \citet{weiss13}, where for $3.6<z<7.5$ we
always expect $^{12}$CO(4--3), $^{12}$CO(5--4) and/or
$^{12}$CO(6--5) in our frequency search range, with other lines
such as C\,{\sc i}(1--0) and H$_2$O(211--202) also present for
some redshifts.

Different approaches were used during the two projects to maximize the
probability of detecting multiple emission lines from each target,
necessary to yield an unambiguous redshift \citep[see discussion
in][]{weiss09}. In W05A, once a single emission line was detected
during an initial sweep of the 3-mm atmospheric window, the remaining
3-mm tunings were skipped and we instead tuned to a higher frequency,
outside the 3-mm band, to search for a higher CO transition, having
used the initial line and/or continuum detection to quantize the
possibilities as well as to improve the photometric redshift
estimate. In X0C6, spectra covering the 3-mm atmospheric window were
obtained for all targets ($\sim$21\,GHz in total), then targets with
emission lines were observed again to search for emission lines at
higher frequencies, in the 2-mm band.  Some of the targets
therefore have less than 21\,GHz of coverage (e.g.\ NGP-190387,
NGP-126191 and G09-59393); others have coverage larger than 21\,GHz
(NGP-284357, NGP-246114 and G09-81106). Average on-source time per
tuning was 20\,min for W05A, and 120\,min for X0C6, of which 15 and
90\,min remained after flagging, respectively.

Calibration of the data was carried out by using the {\sc gildas}
package\footnote{http://www.iram.fr/IRAMFR/GILDAS}. The typical
resulting r.m.s.\ noise levels were 2.7 and 1.1\,mJy\,beam$^{-1}$
in 100-\kms\ channels for data taken in W05A and X0C6. Calibrated
visibilities were converted into FITS format for export, then
into MS format to be imaged by {\sc casa} \citep{mcmullin07}. The
average synthesized beam size was 5\arcsec {\sc fwhm} during both
runs, with considerable diversity in beam shape due to the
relatively short tracks.

To study any 3-mm continuum emission from our targets, we
integrated our data over all observed frequencies, imaging with
the {\sc clean} task in {\sc casa}, with a map size of
$1^{\prime}\times 1^{\prime}$, sufficient to cover the 3-mm
primary beam.

\subsection{ALMA 3-mm spectral scans}

Four ultrared galaxies were observed using ALMA (see
Table~\ref{target}), with five separate tunings to cover the 3-mm
window (Program ID: 2013.1.00499.S; PI: A.~Conley). Data were
acquired during 2014 July 02--03 and August 28, with typically
8.6--9.7\,min spent on-source for each tuning, in addition to
20\,min of calibration -- pointing, phase, flux density (Neptune)
and bandpass.

Data were calibrated using the ALMA pipeline, with only minor flagging
required. Calibrated data were imaged using {\sc clean} within {\sc
  casa}, using the natural weighting scheme to maximize sensitivity.

The resulting r.m.s.\ noise levels ranged between 0.73 and
0.80\,mJy\,beam$^{-1}$ in channels binned to 100\,\kms. Because the
observations were carried out on several different dates, with
different antenna configurations, at frequencies ranging from 84 to
115\,GHz, the resulting synthesized beamsizes varied between 0.6 and
1.2$''$ {\sc fwhm}.

As with our NOEMA data, 3-mm continuum images were created using
all the available data, with a map size of
$1^{\prime}\times 1^{\prime}$.

\subsection{NOEMA 1.3-mm continuum observations}
\label{noema}

We have also carried out 1.3-mm observations of ten galaxies
lacking continuum detections, and hence accurate positions, in
our earlier 3-mm work. Table~\ref{target} lists those targets
observed during 2015 December (Program ID: W15ET; PI: M.~Krips),
again using the most compact NOEMA configuration, with six
antennas. The typical resulting synthesized beam size was
$\sim 1.5''$ {\sc fwhm}. Calibration was accomplished following
the standard procedures, using {\sc gildas}, with little need for
significant flagging. The average time spent on-source was
25\,min, yielding typical noise level of 0.47\,mJy\,beam$^{-1}$.

We also use data from an earlier programme which observed another five
of our targets -- G09-81106, G09-83808, NGP-101333, NGP-126191 and
NGP-246114 -- taken during 2013 in the compact 6C configuration, with
a typical resulting synthesized beam size of $1.0''\times 1.3''$ {\sc
  fwhm}, the major axis at a position angle of 25$^\circ$ (Program ID:
W0BD; Co-PIs: F.~Bertoldi, I.~Perez-Fournon).

\section{Results}
\label{results}

If detecting faint line emission from distant galaxies is
challenging, doing so in the absence of an accurate position is
considerably more so.  For this reason, our first step is to
explore the 3-mm continuum images described in \S\ref{obs},
hoping that thermal dust emission from our luminous, dusty
starbursts will betray the precise position of our targets.

\subsection{Continuum emission}
\label{continuum}

To determine the significance of any continuum emission, we measured
the r.m.s.\ noise level of the maps, and then created the
signal-to-noise ratio (SNR) images shown in Fig.~\ref{detcont1}.
 
All four sources observed at 3\,mm with ALMA are clearly detected in
continuum, at $>8\sigma$ significance.

For the objects observed at 3\,mm with NOEMA, the sensitivity is much
reduced compared to ALMA, so we begin by overlaying the 3-mm continuum
images with contours from the deep SCUBA-2 850-$\mu$m imaging of
\citet{ivison16}, where the unsmoothed {\sc fwhm} of the SCUBA-2
images is around 13$''$, and the r.m.s.\ pointing accuracy of the JCMT
for a single visit to a target is $\sim 2$--3$''$.

We then searched for faint 3-mm continuum sources coincident with
SCUBA-2 850-$\mu$m emission, finding eight plausible sites.  We
discount the faint 3-mm emission seen towards G09-59393, favoring
the 1.3-mm position a few arcsec to the east, which is
considerably more significant.  The most dubious of the others is
NGP-113609, although the close proximity of the 3-mm peak to the
SCUBA-2 850-$\mu$m emission lends extra confidence.  NGP-126191
displays $\ge4$-$\sigma$ emission; again, the near-coincidence
with 850-$\mu$m and/or 1.3-mm emission gives additional
confidence.  For the five remaining sources, 3-mm continuum
emission was detected at $>5\sigma$.

Of the targets observed in continuum at 1.3\,mm using NOEMA, we were
able to measure positions and flux densities for 13 of 15.

The flux densities and coordinates of all these continuum detections
are quoted in Tables~\ref{mescont} and \ref{position}, respectively,
corrected for primary beam attenuation, including the small number of
tentative examples (which are marked as such).  The contribution from
emission lines to the continuum flux density is negligible, as we
shall see in what follows.

It is worth noting here that none of the ultrared galaxies observed in
1.3- or 3-mm continuum are revealed as doubles, as would be expected
in the simulations of \citet{bethermin17}, though it remains possible
that some or all of the $\sim 20$\% of targets that remain undetected
in continuum have been pushed below our interfeometric detection
threshold by multiplicity.

\begin{table*}
	\begin{center}
	\caption{Continuum flux-density measurements and redshifts,
	photometric and/or spectroscopic.}
	\label{mescont}
	\begin{tabular}{lccccccccc} 
		\hline\hline
		Nickname & $S_{250}$ &
		$S_{350}$ & $S_{500}$ &
		$S_{850}$ & $S_{\rm 1.3\,mm}^{\rm a}$ & $S_{\rm 3\,mm}^{\rm a}$ & $z_{{\rm phot}}^{\rm b}$ & $z_{{\rm spec}}$ & Reference \\
		\hline
		SGP-196076 &$28.6\pm7.3$&$28.6\pm8.2$&$46.2\pm8.6$&$32.5\pm9.8$&--&$0.41\pm0.03$&$4.51^{+0.47}_{-0.39}$&$4.425\pm0.001$&--\\
		SGP-261206 &$22.6\pm6.3$&$45.2\pm8.0$&$59.4\pm8.4$&$56.9\pm8.9$&--&$0.38\pm0.02$&$5.03^{+0.58}_{-0.47}$&$4.242\pm0.001$&--\\
		SGP-354388 &$26.6\pm8.0$&$39.8\pm8.9$&$53.5\pm9.8$&$39.9\pm4.7$ &--&$0.35\pm0.02$&$5.35^{+0.56}_{-0.52}$&$4.002\pm0.001$&\citet{oteo17grh}\\
		SGP-32338 &$16.0\pm7.1$&$33.2\pm8.0$&$63.7\pm8.7$&$27.9\pm9.4$&--&$0.21\pm0.02$&$3.93^{+0.26}_{-0.24}$&--&--\\
		G09-59393  &$24.1\pm7.0$ &$43.8\pm8.3$ &$46.8\pm8.6$ &$27.7\pm5.6$ &$4.0\pm0.6$&-- &$3.70^{+0.35}_{-0.26}$ &--&--\\
		G09-81106  &$14.0\pm 6.0$ &$30.9\pm8.2$ &$47.5\pm 8.8$ &$37.4\pm 11.4$ & $9.7\pm 1.3$ &$0.24\pm0.04$ &$ 4.98^{+0.13}_{-0.73}$&$4.531\pm0.001$&--\\
		G09-83808&$9.7\pm5.4$ &$24.6\pm 7.9$ &$44.0\pm 8.2$ &$36.2\pm9.1$& $19.4\pm 2.0$ &$0.66\pm 0.12$ &$5.66^{+0.06}_{-0.74}$ & $6.027\pm 0.001$&\citet{zavala17}\\
		G09-62610 &$18.6\pm 5.4$ &$37.3\pm7.4$ &$44.3\pm7.8$ &$23.1\pm9.0$&$5.2\pm0.8$&$\le 0.18$ &$3.70^{+0.44}_{-0.26}$ &--&--\\
		G15-26675 &$26.8\pm6.3$ &$57.2\pm 7.4$ &$61.4\pm 7.7$ &$36.6\pm 10.3$ &--&$\le 0.33$ &$4.36^{+0.25}_{-0.21}$ &--&--\\
		G15-82684 &$17.3\pm6.4$ &$38.5\pm 8.1$ &$43.2\pm 8.8$ &$15.3\pm 8.2$ &$\le 1.6$&$\le 0.36$ &$3.65^{+0.38}_{-0.25}$ &--&--\\
		NGP-206987 &$24.1\pm7.1$ &$39.2\pm8.2$ &$50.1\pm8.7$ &$17.5\pm6.5$ &$9.2\pm1.8$&$\le 0.32$ &$4.07^{+0.06}_{-0.60}$ &--&--\\
		NGP-111912 &$25.2\pm6.5$ &$41.5\pm 7.6$ &$50.2\pm8.0$ &$8.8\pm 6.7$ &$4.7\pm0.9$&$\le0.26$ &$3.27^{+0.36}_{-0.26}$ &--&--\\
		NGP-136156 &$29.3\pm 7.4$ &$41.9\pm8.3$ &$57.5\pm 9.2$ &$29.7\pm 4.6$ &$3.1\pm0.8$&$\le 0.25$ &$3.95^{+0.06}_{-0.57}$ &--&--\\
		NGP-126191 &$24.5\pm6.4$ &$31.3\pm7.7$ &$43.7\pm 8.2$ &$37.2\pm7.5$ & $12.3\pm 1.7$ &$0.30\pm0.11$ $^{\rm c}$ &$4.33^{+0.45}_{-0.46}$ &--&--\\
		NGP-284357 &$12.6\pm 5.3$ &$20.4\pm 7.8$ &$42.4\pm 8.3$ &$27.4\pm9.9$ &--&$0.62\pm0.03$ &$4.99^{+0.44}_{-0.45}$ &$4.894\pm0.003$&--\\
		NGP-190387 &$25.2\pm 7.2$ &$41.9\pm 8.0$ &$63.3\pm 8.8$ &$33.4\pm 8.0$ &$12.2\pm1.2$&$0.84\pm 0.14$ &$4.36^{+0.37}_{-0.26}$ &$4.420\pm0.001$&--\\
		NGP-113609 &$29.4\pm7.3$ &$50.1\pm8.0$ &$63.5\pm8.6$ &$12.5\pm6.2$ &$13.0\pm2.3$&$\le0.26 $ &$3.43^{+0.34}_{-0.20}$ &--&--\\
		NGP-252305 &$15.3\pm 6.1$ &$27.7\pm8.1$ &$40.0\pm9.4$ &$23.5\pm 7.6$ &$6.5\pm0.7$&$\le 0.29$ &$4.34^{+0.43}_{-0.38}$ &--&--\\
		NGP-63663 &$30.6\pm6.8$ &$53.5\pm7.8$ &$50.1\pm8.1$ &$7.9\pm8.3$ &$\le 1.3$&$\le0.24 $ &$3.08^{+0.23}_{-0.22}$ &--&--\\
		NGP-246114 &$17.3\pm6.5$ &$30.4\pm 8.1$ &$33.9\pm8.5$ &$32.4\pm8.2$&$8.0\pm1.5$&$0.42\pm0.06$ &$4.35^{+0.53}_{-0.46}$ &$3.847\pm0.002$&--\\
		NGP-101333 &$32.4\pm7.5$ &$46.5\pm8.2$ &$52.8\pm 9.0$ &$17.6\pm8.2$&$10.8\pm1.3$&$\le 0.25$ &$3.53^{+0.34}_{-0.27}$ &--&--\\
		\hline
	\end{tabular}
	\end{center}
	\begin{flushleft}
	$\rm^{a}$ Measured flux density, or $3\sigma$ upper
		   limit. Stated errors exceed the local r.m.s.\ in the relevant image, since they
		   reflect all uncertainties, including source size.\\
	$\rm^{b}$ Photometric redshift estimated by template SED fits to 250-, 350-, 500-, and 850- or 870-$\mu$m flux densities \citep{ivison16}.\\
	$\rm^{c}$ Tentative detection only.
	\end{flushleft}
\end{table*}

\begin{table}
	\begin{center}
	\caption{Precise J2000 positions of ultrared galaxies.}
	\label{position}
	\begin{tabular}{lcc} 
		\hline\hline
		Nickname & R.A. & Dec.\\
		\hline
		 SGP-196076$^{\rm b}$ (aka SGP-38326)&00:03:07.22&$-$33:02:50.9\\
		 SGP-261206$^{\rm b}$ &00:06:07.54&$-$32:26:39.9\\
		 SGP-354388$^{\rm b}$ (aka GRH)&00:42:23.52&$-$33:43:23.5\\
		 SGP-32338$^{\rm b}$ &01:07:41.00&$-$28:27:09.4\\
		 G09-59393$^{\rm a}$ &08:41:13.42&$-$00:41:11.7\\
		 G09-81106$^{\rm b}$ &08:49:36.82&+00:14:54.7\\
		 G09-83808$^{\rm a,b}$ &09:00:45.79&+00:41:22.9\\
		 G09-62610$^{\rm a}$ &09:09:25.18&+01:55:43.7\\
		 G15-26675 &--&--\\
		 G15-82684 &--&--\\
		 NGP-206987$^{\rm a}$ &12:54:40.67&+26:49:29.6\\
		 NGP-111912$^{\rm a}$ &13:08:24.04&+25:45:17.9\\
		 NGP-136156$^{\rm a}$ &13:26:27.57&+33:56:35.5\\
		 NGP-126191$^{\rm a}$ &13:32:17.76&+34:39:47.5\\
		 NGP-126191$^{\rm d}$ &13:32:17.82&+34:39:50.5\\
		 NGP-284357$^{\rm b}$ &13:32:51.73&+33:23:42.8\\
		 NGP-190387$^{\rm a}$ &13:33:37.47&+24:15:39.3\\
		 NGP-113609$^{\rm a,c}$ &13:38:36.65&+27:32:53.5\\
		 NGP-252305$^{\rm a}$ &13:39:19.27&+24:50:59.4\\
		 NGP-63663 &--&--\\
		 NGP-246114$^{\rm a,b}$ &13:41:14.09&+33:59:38.2\\
		 NGP-101333$^{\rm a}$ &13:41:19.36&+34:13:46.5\\
		\hline
	\end{tabular}
	\end{center}
	$\rm^{a}$ Position determined via 1.3-mm continuum. \\
	$\rm^{b}$ Position determined via 3-mm continuum. \\
	$\rm^{c}$ Tentative continuum detection. \\
	$\rm^{d}$ Position determined via emission line.
\end{table}

\subsection{Searching for emission lines}

To determine reliable, unambiguous redshifts for a DSFG, we must
detect two or more emission lines.  Ideally we must extract their
spectra at known positions, typically betrayed by interferometric
continuum detection in the cases of DSFGs, thereby maximising the
significance of any line detections. If we extract spectra blindly, we
must correct our statistics for the number of independent sightlines
explored.  Here, our known positions come from the 1.3- and 3-mm
continuum imaging with NOEMA and ALMA, as described in
\S\ref{continuum}; for the 17 sources with reliable coordinates
(Table~\ref{position}), we extracted spectra at the precise positions
of the corresponding continuum detections. 

In the three cases where we have no continuum detection at either 1.3
or 3\,mm, tagged as such in Table~\ref{position}, we searched blindly
for emission lines in data cubes that had not been corrected for the
primary beam response.  We convolved these data cubes along their
frequency axis with box-car kernels of width 3, 4 and 7 channels,
corresponding to velocity widths of $\approx 200$--500\,km\,s$^{-1}$,
typical for DSFG emission lines \citep{bothwell13}.  For each
convolved cube we created a SNR cube, then searched for peaks above
$5\sigma$, where the significance of detections at this stage have not
been corrected for the number of independent sightlines we have
explored.  We also performed the same blind line-search procedure on
continuum-detected sources to look for any additional line emission.
Only known lines were recovered.

As a result of these emission-line searches, we detected multiple (two
or more) emission lines from seven of our targets, one of these
following the detection of three lines by \citet{zavala17}, as well as
single emission lines from four targets, where more lines have been
detected subsequently in one case \citep{oteo17grh}.  We thus report
the first eight robust, accurate, unambiguous redshifts for faint,
largely unlensed and thus intrinsically very luminous starbursts.

For all the detected emission lines, we have fitted single-component
Gaussians, measuring the frequency of the line center, and its {\sc
  fwhm}.  Continuum emission was subtracted with the UVCONTSUB task in
{\sc casa}, using all available channels except those close to
emission lines.  The flux of each emission line has been measured with
the {\sc casa} IMFIT task, from the zeroth moment map (created by
integrating along the frequency axis across the emission line).  There
are no significant discrepencies between these values and those found
from the Gaussian fits.  The measured properties of the emission lines
are summarized in Table~\ref{emission}.

\subsection{Unambiguous redshifts via detection of multiple emission lines}

\begin{figure}
\centerline{\includegraphics[width=3.5in]{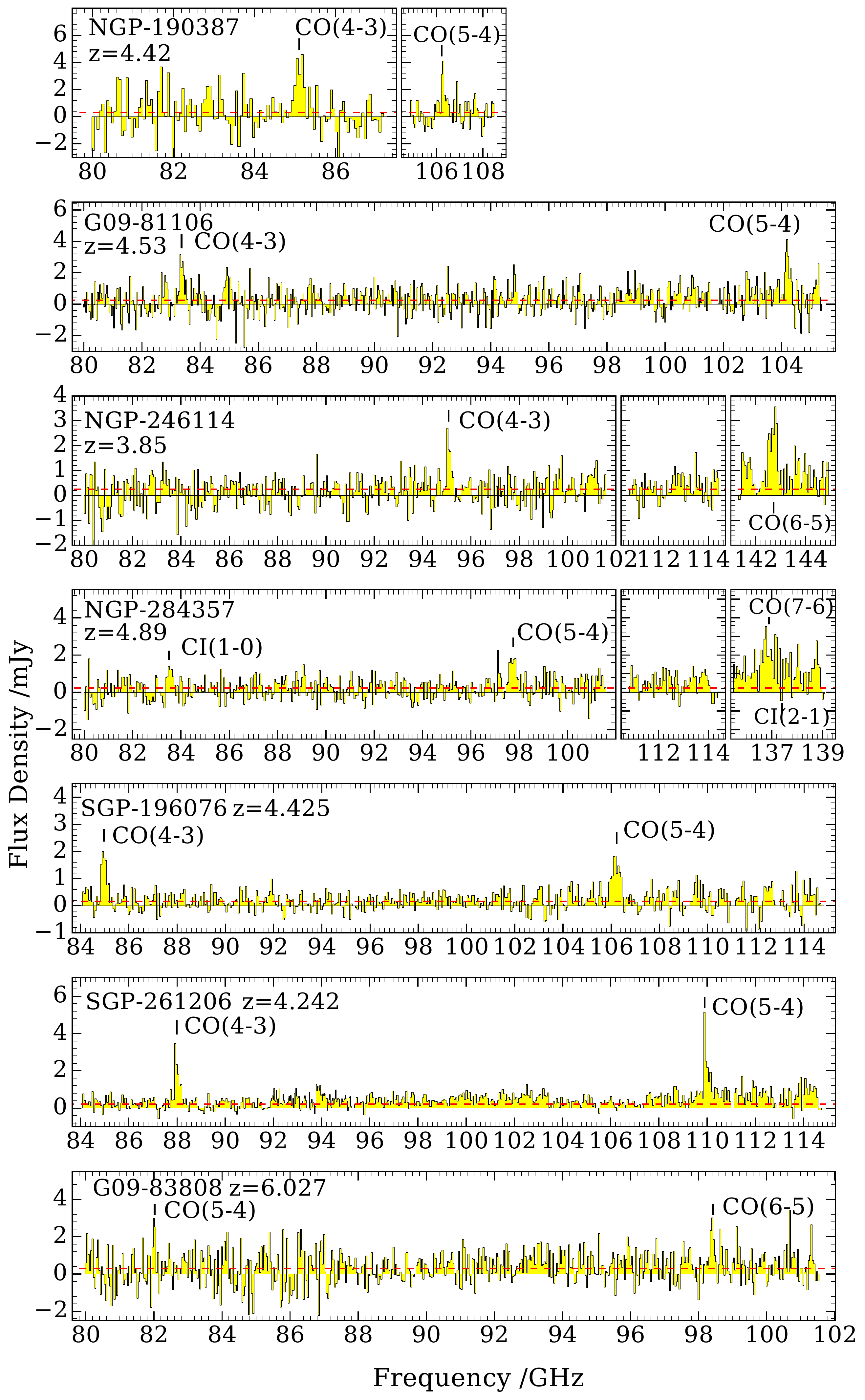}}
\caption{3-mm spectra of targets with clear multiple line detections,
  thus yielding unambiguous redshifts, extracted at the positions
  where 3-mm continuum is seen.  The spectra have been binned to 250,
  150, 150, 200, 100, 100 and 100\,\kms, from top to bottom,
  respectively. Red dashed lines show the median continuum flux
  density value calculated across the full frequency range, excluding
  $\pm 0.5$\,GHz around the emission lines. All but two of the
  spectroscopic redshifts agree well with the photometric estimates of
  \citet{ivison16} -- see Table~\ref{compare}. NGP-284357 shows
  C\,{\sc i}(1--0) emission at the expected position, 83.6\,GHz,
  however the line falls between WideX tunings, meaning the line
  properties are difficult to measure. C\,{\sc i}(2--1) is located at
  $\sim 137.41$, blended alongside $^{12}$CO(7--6). Like many DSFGs,
  SGP-196076 comprises merging galaxies \citep{oteo16}, and here we
  show the combined spectrum of the two most luminous components.}
\label{doublee1}
\end{figure}

We detect multiple emission lines towards seven of our targets, such
that the redshifts of these sources and species/transitions of the
emission lines are confirmed unambiguously.  The properties of those
sources are discussed in \S\ref{sedfit}.

\begin{figure}
\centerline{\includegraphics[width=3.5in]{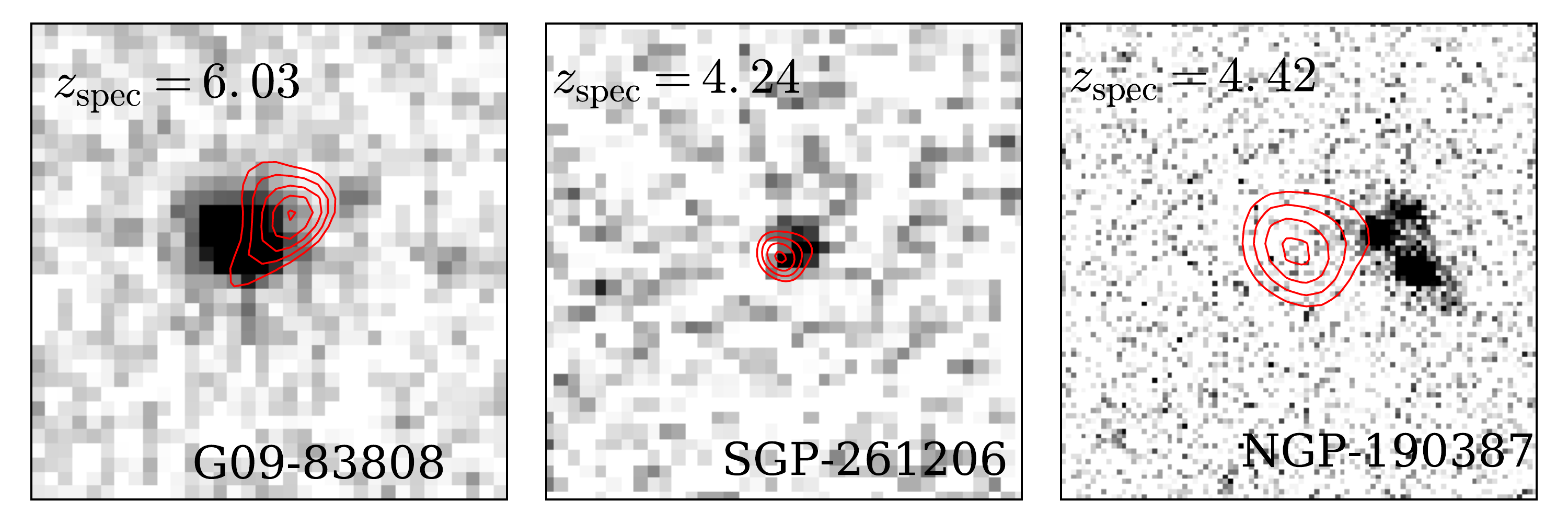}}
 \caption{Near-IR ($K$-band) $11''\times 11''$ images of G09-83808 and
   SGP-261206 from VIKING survey \citep[][]{edge13}, and NGP-190387
   from Gemini/NIRI (programme GN-2016A-FT-32).  Red contours
   represent the 1.3-mm (G09-83808, NGP-190387) and 3-mm (SGP-261206)
   continuum emission, at $5,6,7,8,9\times\sigma$.  Near-IR emission
   is seen clearly, coincident or near-coincident with the 1.3-mm or
   3-mm continuum emission. At the depth of the VIKING observations
   (limiting magnitude, $K_{\rm AB}\simeq 21.2$), we do not expect to
   detect dust-obscured distant galaxies. The sources coincident with
   G09-83808 and SGP-261206 show, therefore, that these two DSFGs are
   amplified gravitationally by the foreground galaxies seen in the
   near-IR images. The galaxy in the foreground of SGP-83808 has
   $z_{\rm spec} = 0.778$, obtained using X-shooter on the VLT
   (Fig.~\ref{figxshooter}), and is magnified by $8.2\pm0.3$, a factor
   determined using high-resolution ALMA continuum imaging
   \citep[see][]{oteo17morph}.  The three galaxies revealed by the
   deeper Gemini imaging of NGP-190387 are considerably fainter, with
   $K_{\rm AB}=21.0, 22.3, 21.4$, yet they are not coincident with the
   $z=4.42$ DSFG and likely constitute a foreground lensing group.}
\label{nir}
\end{figure}

For NGP-190387, two emission lines are detected at 85.10 and at
106.23\,GHz (see Fig.~\ref{doublee1}), CO(4--3) and CO(5--4) at
$z=4.420$ ($z=4.418$ and 4.425, respectively, for the two
lines). NGP-190387 lies close to a group of three faint
($K_{\rm AB}\approx 21$--22) galaxies, likely at $z\gs1$, revealed by
NIRI on the 8-m Gemini North telescope (Fig.~\ref{nir}), which amplify
the DSFG gravitationally by a factor we cannot constrain meaningfully
at the present time.

Towards G09-81106 we have detected two emission lines, CO(4--3)
and CO(5--4), at 83.36 and 104.19\,GHz (Fig.~\ref{doublee1}),
both at $z=4.531$.  There is no suggestion of gravitational
lensing for G09-81106, either via the presence of unusually
bright near-IR galaxies in the field, or via its submm morphology
as seen in high-resolution ALMA continuum imaging
\citep{oteo17morph}.

\begin{figure}
\centerline{\includegraphics[width=3.5in]{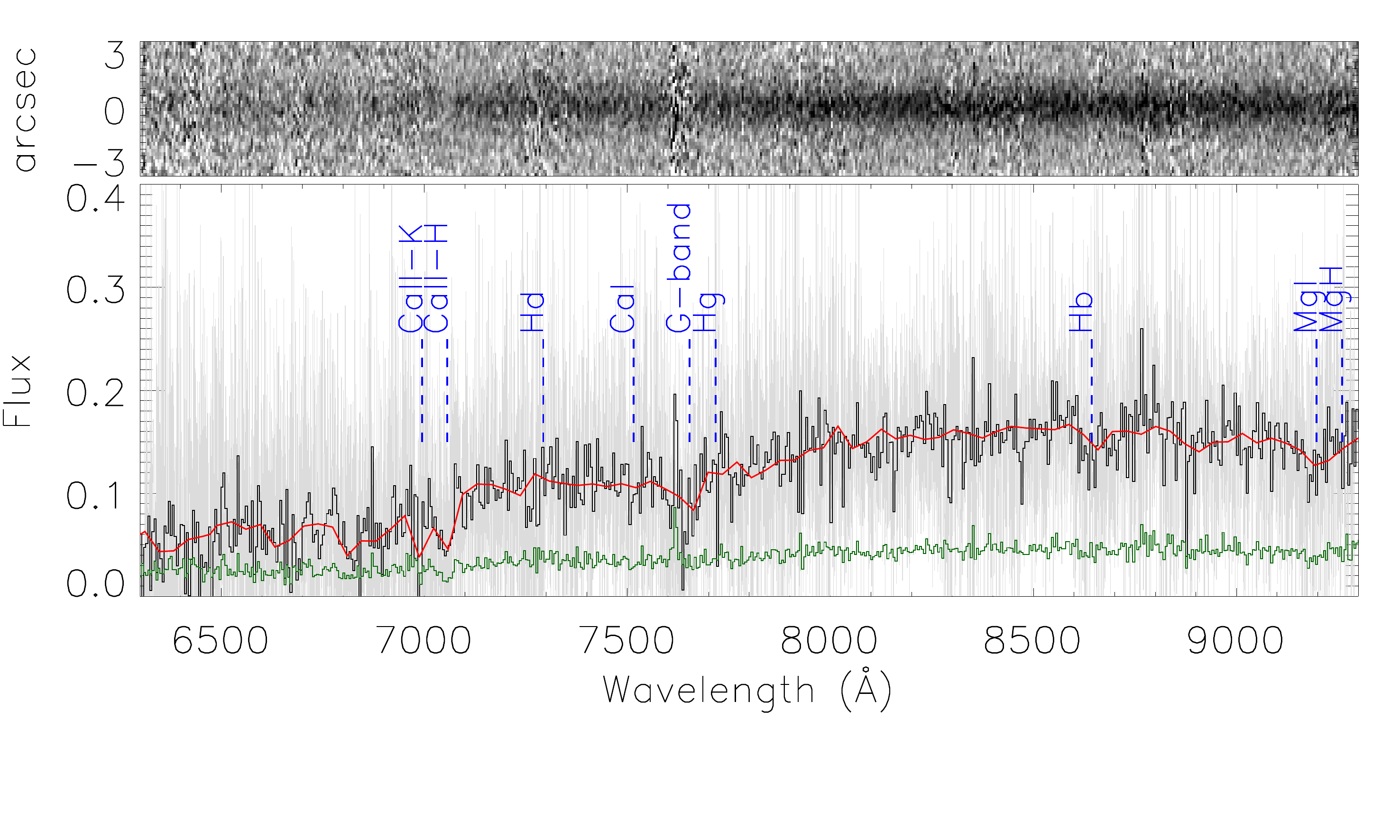}}

\caption{{\it Top:} Spectrum of the lensing galaxy in the foreground
  of SGP-83808, as seen in the optical arm of VLT/X-shooter after
  integrating for $4\times 20$\,min on 2013 March 17.  {\it Below:}
  The 1-D spectrum, in grey, extracted using a simple box-car
  summation [090.A-0891(A), PI: Christensen]. To identify the lens
  redshift we binned these data heavily
  \citep[see][]{modigliani10,christensen12, kausch14}, as illustrated
  in black, with the error spectrum shown in green, then fitted the
  redshift by $\chi^2$ minimisation using a 5-Gyr stellar population
  model (shown in red) from \citet{bc03}.  We find
  $z_{\rm spec}=0.778\pm0.006$. Strong absorption features are
  labelled. The flux units are $10^{-17}$\,erg\,s$^{-1}$\,cm$^{-2}$\,\AA$^{-1}$.}
\label{figxshooter}
\end{figure}

Towards G09-83808 we have detected two faint emission lines, at 82.02
and 98.39\,GHz (Fig.~\ref{doublee1}), corresponding to CO(5--4) and
CO(6--5) at $z=6.026\pm0.001$ and $6.028\pm0.001$, respectively, so an
average of $6.027\pm0.001$.  These lines were also noted by
\citet{zavala17} in a spectrum obtained using the Large Millimeter
Telescope.  G09-83808 is near-coincident with a foreground galaxy,
seen clearly in near-IR imaging from the VIKING survey \citep[see
Fig.~\ref{nir} --][]{edge13}, indicative of gravitationally lensing.
This foreground galaxy has a spectroscopic redshift of 0.778, obtained
using X-shooter on the 8-m Very Large Telescope (see
Fig.~\ref{figxshooter}).  A lens model based on the morphology
determined by high-resolution ALMA continuum imaging predicts a
gravitional amplification of $8.2\pm 0.3$ \citep{oteo17morph}.

For NGP-284357, we find at least two emission lines, at 97.75 and
136.9\,GHz (Fig.~\ref{doublee1}). If these are CO(5--4) and CO(7--6),
the redshifts are 4.895 and 4.892, respectively, so an average of
$z=4.894$.  At this redshift, the fine-structure lines of neutral
carbon are expected at 83.54 and 137.41\,GHz, respectively, and we see
strong hints of corresponding emission -- a discrete feature where
C\,{\sc i}(1--0) is expected, and C\,{\sc i}(2--1) appears to be
broadening the CO(7--6) line.

In the case of NGP-246114, two emission lines are detected at 95.08
and 142.71\,GHz (Fig.~\ref{doublee1}), which must be CO(4--3) and
CO(6--5) at $z=3.849$ and 3.845, respectively, so an average of
$z=3.847$.

SGP-196076 has been studied in detail by \citet{oteo16}, who referred
to the galaxy as SGP-38326, an {\it H}-ATLAS nomenclature pre-dating
\citet{ivison16}; for further details we refer the readers to that
paper.  Summarising the main results obtained from our 3-mm spectral
scans of SGP-196076: we have detected the CO(5--4) and CO(4--3)
transitions at 84.97 and 106.19\,GHz, so at $z=4.425$
(Fig.~\ref{doublee1}). C\,{\sc i} is also seen, at low
significance. Both the continuum emission, from dust, and the CO(5--4)
line emission, indicate clearly that SGP-196076 comprises multiple
($\ge 3$) components, with the star formation in each one presumably
triggered by their close proximity -- a ongoing merger or strong
interaction.  \citet{oteo16} explored the velocity field of the two
largest components, via their CO and [C\,{\sc ii}] emission, finding
ordered disk-like rotation.

SGP-261206 displays emission lines at 87.95 and 109.01\,GHz
(Fig.~\ref{doublee1}), CO(4--3) and CO(5--4) at
$z=4.242\pm 0.001$, around $1.7\sigma$ below its photometric
redshift. C\,{\sc i} is also seen, at low significance. Very
dust-obscured, distant galaxies should not be coincident with
near-IR sources at the depth of our available imaging, unless
those near-IR sources are gravitationally lensing the dusty
galaxy.  However, the $K$-band image\footnote{Gravitational
  lensing is found likely for three of the galaxies in this
  sample, as revealed by $K$-band imaging -- see Fig.~\ref{nir};
  the rest are devoid of close near-IR counterparts, though the
  depth of the available near-IR imaging does not exclude the
  possibility of distant ($z\gs 1$) lenses.} of SGP-261206 shown
in Fig.~\ref{nir}, from VIKING \citep{edge13}, contains a clear
$K$-band counterpart, coincident with the dust emission. This
suggests that SGP-261206 is gravitationally lensed by the
foreground galaxy detected in the near-IR image, a hint confirmed
by high-resolution ALMA imaging \citep{oteo17morph}.

\begin{table}
	\begin{center}
	\caption{Measured properties of the detected emission lines.}
	\label{emission}
	\begin{tabular}{ccccc} 
		\hline\hline
		Nickname&Transition&$\nu_{\rm line}$
		&Flux$^{\rm a}$&{\sc fwhm}$^{\rm b}$\\
		&&/GHz&/${\rm Jy\,km\,s^{-1}}$
		&/${\rm km\,s^{-1}}$\\
		\hline
SGP-196076$^{\rm d}$&CO(4--3)&$84.97\pm0.01$&$3.18\pm0.34$&$1080\pm90$\\
         &CO(5--4)&$106.19\pm0.02$&$1.22\pm0.12$&$1280\pm80$\\
SGP-261206&CO(4--3)&$87.95\pm0.01$&$2.12\pm0.33$&$440\pm40$\\
         &CO(5--4)&$109.01\pm0.01$&$2.94\pm0.33$&$440\pm30$\\
  SGP-354388&C\,{\sc i}(1--0)&$98.39\pm0.01$&$0.97\pm0.22$&$700\pm180$\\
  SGP-32338&CO(5--4)$^{\rm c}$&$100.07\pm0.01$&$1.70\pm0.20$&$630\pm80$\\
  G09-81106&CO(4--3)&$83.36\pm0.01$&$1.27\pm0.21$&$570\pm	130$\\
        &CO(5--4)&$104.19\pm0.01$&$1.56\pm0.33$&$470\pm100$\\
  G09-83808&CO(5--4) &82.02$\pm$0.02&$0.92\pm0.30$&$240\pm 100$\\
        &CO(6--5)&$98.39\pm0.01$&$0.87\pm0.24$&$360\pm 110$\\
NGP-111912&CO(4--3)$^{\rm c}$ &$95.15\pm0.04$&$2.04\pm0.79$&$440\pm200$ \\
NGP-126191&CO(5--4)$^{\rm c}$ &$85.77\pm0.02$&$3.19\pm0.88$&$570\pm180$\\
  NGP-284357&CO(6--5)&$97.75\pm0.03$&$2.37\pm0.52$&$680\pm180$\\
        &CO(7--6)&$136.90\pm0.09$&$2.70\pm0.68$&$420\pm150$\\
NGP-190387&CO(4--3)&$85.10\pm0.02$&$2.52\pm0.67$&$670\pm250$ \\
        &CO(5--4)&$106.23\pm0.02$&$2.69\pm0.71$&$440\pm150$ \\
NGP-246114&CO(4--3)&$95.08\pm0.02$&$1.36\pm0.19$&$550\pm150$\\
        &CO(6--5)&$142.71\pm0.03$&$1.60\pm0.32$&$660\pm140$\\
		\hline
	\end{tabular}
	\end{center}
	$\rm^{a}$ Measured via 2-D Gaussian fit to zeroth moment image, after continuum subtraction.\\
	$\rm^{b}$ {\sc fwhm} calculated via Gaussian fit to spectrum with 100-\kms\ spectral resolution.\\
	$\rm^{c}$ Most probable CO transition, based on the photometric redshift estimate from \citet{ivison16}.\\
	$\rm^{d}$ Properties measured by combining all components.
\end{table}

\subsubsection{CO line ratios}

Our spectra allow us to determine line luminosity ratios for those
galaxies for which multiple lines were detected, typically anchored to
$^{12}$CO $J=5$--4.

In Table~\ref{ratio} we list the CO line luminosity ratios (i.e.\
$L^{\prime}_{\text{CO(i - i-1)}}/L^{\prime}_{\text{CO(j - j-1)}}$)
which we find are consistent with the average values found for SMGs by
\citet{bothwell13}.

\begin{table}
	\begin{center}
	\caption{Line luminosity ratios of CO lines }
	\label{ratio}
	\begin{tabular}{lccc} 
		\hline\hline
		Object&CO transitions
		&Line luminosity ratio&Bothwell$^{\rm a}$\\
		\hline
SGP-196076&5--4/4--3&$0.81\pm0.15$&$0.78\pm0.18$\\
SGP-261206&5--4/4--3&$0.90\pm0.17$&$0.78\pm0.18$\\
G09-81106&5--4/4--3&$0.80\pm0.21$&$0.78\pm0.18$\\
G09-83808&6--5/5--4&$0.66\pm0.28$&$0.66\pm0.16$\\
NGP-284357&7--6/5--4&$0.58\pm0.19$&$0.56\pm0.15$\\
NGP-190387&5--4/4--3&$0.69\pm0.26$&$0.78\pm0.18$\\
NGP-246114&6--5/4--3&$0.52\pm0.13$&$0.51\pm0.13$\\
		\hline
	\end{tabular}
	\end{center}
	$\rm^{a}$ Average line luminosity ratio for SMGs from \citet{bothwell13}.
\end{table}

\subsubsection{Rest-frame stacking}

\begin{figure*}
\centerline{\includegraphics[width=7in]{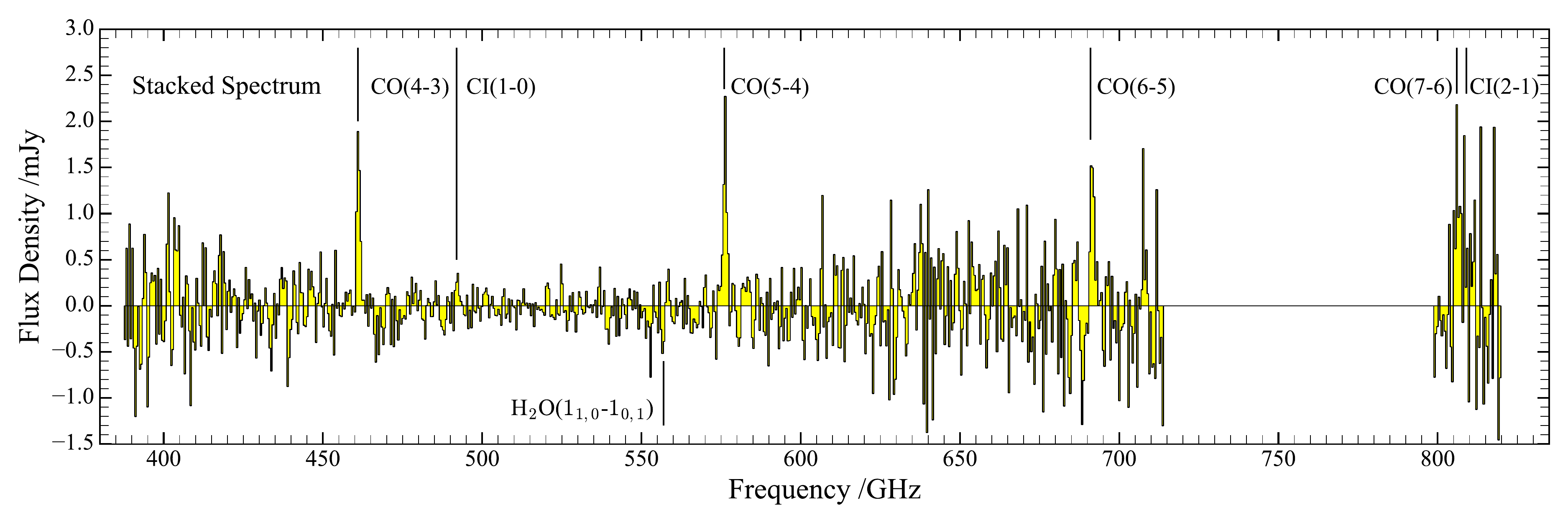}}
\caption{Rest-frame stacked spectrum, noise-weighted and
  continuum-subtracted, of the eight galaxies for which we have
  accurate, unambiguous redshifts, with lines marked, including
  the $^{12}$CO ladder between $J=4$--3 and $J=7$--6, the latter
  broadened by C\,{\sc i}(2--1), as well as weak C\,{\sc i}(1--0)
  line emission.  The position of H$_2$O $1_{1,0}$--$1_{0,1}$ is
  also marked, though the significance of the possible absorption
  line is only $\approx 2.5\sigma$.}
\label{stackfig}
\end{figure*}

For the eight spectra for which we have accurate, unambiguous
redshifts, we can shift the data to the corresponding rest-frame
frequencies and stack them to search for features fainter than
the relatively bright $^{12}$CO lines, following
\citet{spilker15} and \citet{zhang17}.  The resulting stacked
spectrum is shown in Fig.~\ref{stackfig} where we find the
expected $^{12}$CO ladder between $J=4$--3 and $J=7$--6, the
latter broadened by C\,{\sc i}(2--1), as well as weak C\,{\sc
  i}(1--0) line emission.  Absorption due to the collisionally
excited H$_2$O $1_{1,0}$--$1_{0,1}$ ground
transition\footnote{Due to its very high critical density, this
  line is very difficult to excite in emission, but it can be
  seen relatively easily in absorption, where there is strong
  background continuum.  In the cold ISM, water is normally
  frozen out, forming icy mantles on dust grains; detecting this
  transition in absorption suggests water is gaseous, perhaps
  because of turbulence or shock heating.} may be seen, at low
($\approx2.5\sigma$) significance.

\subsubsection{Detection of single emission lines}
\label{singlee}

Towards four of our galaxies, single emission lines were detected,
insufficient to determine the redshift of the source unambiguously, as
the species and/or transition of the emission line is
unknown. However, combining the redshift constraint available by
virtue of far-IR/submm color, often only a handful of strong emission
lines become plausible candidates.

\begin{figure}
\centerline{\includegraphics[width=2.5in]{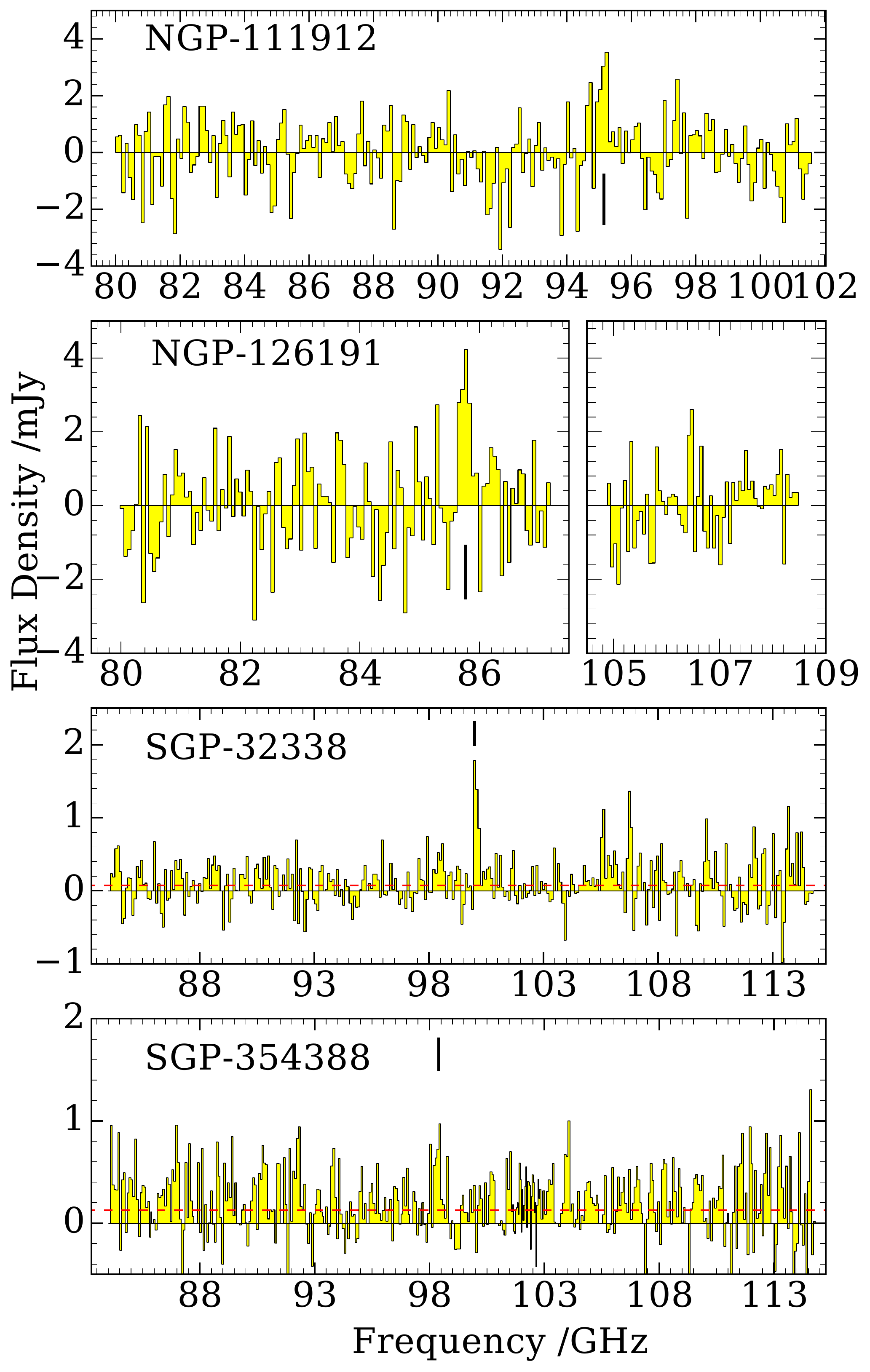}}
\caption{3-mm spectra of NGP-111912, NGP-126191, SGP-32338 and SGP-354388,
  with channels binned to 350, 200, 200, and 350\,\kms,
  respectively. The identification of the emission line and redshift are
  ambiguous for these sources, as no other emission line are detected
  convincingly.  For NGP-111912, the spectrum has been extracted at
  the position of the 1.3-mm continuum; if we assume the emission line
  at 95.15\,GHz is CO(4--3) at $z=3.84$, where the photometric
  redshift estimate is $3.27^{+0.36}_{-0.26}$. For NGP-126191, the
  emission line seen at 85.77\,GHz suggests $z=4.38$, but the
  non-detection of another emission line near 107.1\,GHz makes this
  unlikely. If we instead assume that NGP-126191 lies at $z= 5.71$,
  the emission line at 85.77\,GHz becomes CO(5--4); $z\sim3.03$ or
  $z\sim7.06$ are also feasible. For SGP-32338 \citep[where
  $z_{\rm ph}=4.51^{+0.47}_{-0.39}$][]{ivison16}, the emission line
  detected at 100.07\,GHz could be CO(5--4) at $z=4.70$. We would not
  then expect to detect other lines, despite the wide frequency
  coverage.}
\label{singleefig}
\end{figure}

Towards NGP-126191 we detected a clear emission line at 85.77\,GHz
(Fig.~\ref{singleefig}), with a {\sc fwhm} of $570\pm 180$\,km\,s$^{-1}$.
As outlined earlier, this line emission is $\approx 3''$ from weak
3-mm continuum emission, which may be spurious, or may be from a
companion, or the dust emission may be slightly displaced from the
line emission -- a relatively common finding amongst DSFGs
\citep[e.g.][]{ivison10leblob, fu13, dye15, spilker15, oteo16}.  With
a far-IR/submm photometric redshift of 4.9, the most likely
identification for this emission line is $^{12}$CO(4--3) at $z=4.38$;
however, $^{12}$CO(5-4) would then be expected at 107.1\,GHz, with a
similar significance given the typical spectral-line energy
distributions of DSFGs, and such a line is not detected
(Fig.~\ref{singleefig}).  $^{12}$CO(3--2) and $^{12}$CO(5--4) are the
other most likely possibilities, at $z=3.03$ and $z=5.71$.

Towards NGP-111912 we detected a weak emission line at 95.15\,GHz
(Fig.~\ref{singleefig}), with a {\sc fwhm} of $440\pm 200$\,km\,s$^{-1}$.
The line emission is coincident spatially with 1.3-mm continuum
emission (Fig.~\ref{detcont1}).  With a photometric redshift estimate
of $3.28^{+0.36}_{-0.26}$, the emission line may be $^{12}$CO(4--3) at
$z=3.84$, in which case we would not expect any other lines in our
current frequency coverage, consistent with our data.

SGP-32338 is a similar case: we detected an emission line at
101.07\,GHz, with a {\sc fwhm} of $630\pm 80$\,km\,s$^{-1}$.  The line
emission is again coincident with its 3-mm continuum emission
(Fig.~\ref{detcont1}).  The photometric redshift,
$4.51^{+0.47}_{-0.39}$, makes $^{12}$CO(5--4) at $z=4.70$ the most
likely candidate emission line.  Because the line lies close to the
center of the spectral coverage, we would not then expect to detect
any other lines, despite the high sensitivity and the wide frequency
range available.

Follow-up observations are required to determine unambiguous redshifts
for these three galaxies.

In SGP-354388, dubbed the `Great Red Hope' because it is amongst
the reddest galaxies seen by {\it Herschel}, our ALMA spectrum
reveals a line at 98.34\,GHz, coincident with 3-mm continuum
emission.  Extensive further follow-up observations of
SGP-354388, reported by \citet{oteo17grh}, confirm that the line
at 98.34\,GHz is, in fact, the C\,{\sc i}(1--0) transition at
$z=4.002\pm0.001$, a rare $\approx2\sigma$ deviation from the
photometric redshift which can be attributed at least partially
to dusty galaxies surrounding SGP-354388, at the same redshift,
which contaminate the flux densities measured at $\ge500$\,$\mu$m
by SPIRE and LABOCA \citep{ivison16}.

\subsubsection{Galaxies where no emission lines are detected}
\label{nolines}

\begin{figure}
\centerline{\includegraphics[width=3.2in]{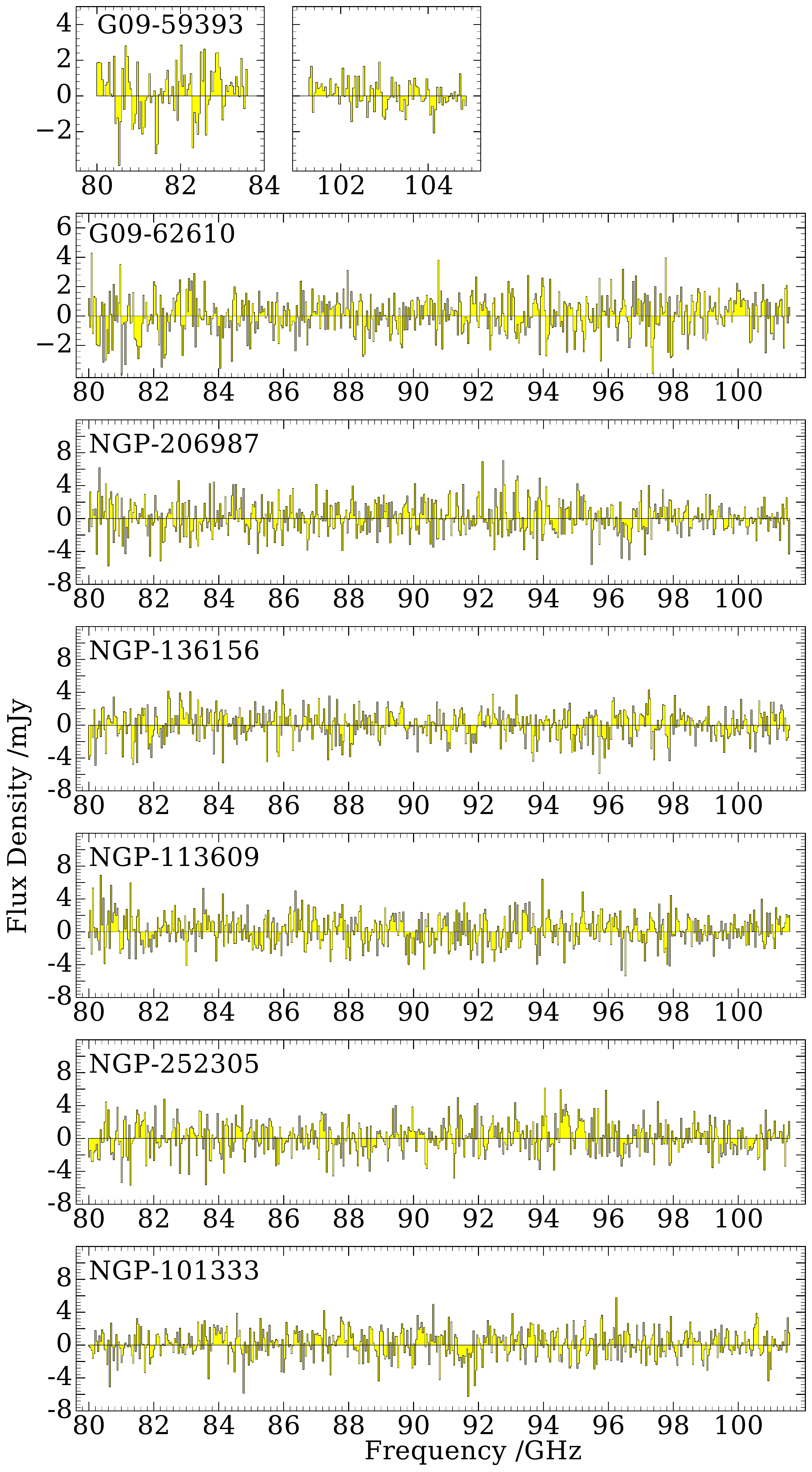}}
\caption{3-mm spectra of G09-59393, G09-62619, NGP-206987,
  NGP-136156, NGP-113609 and NGP-252305, with channels binned to
  200\,\kms, all cases where we have secure positions via
  continuum detections at 1.3 and/or 3\,mm, but where there is no
  strong evidence of line emission.  The median \lir\ for these
  galaxies, based on their photometric redshifts, is
  0.3--0.4\,dex below that of the galaxies with significant line
  emission, which goes some way towards explaining why we have
  detected no line emission in these cases.}
\label{nolinesfig}
\end{figure}

In our remaining spectral scans, regardless of whether or not we
have secure positions via continuum detections, we have found no
compelling evidence of line emission (Fig.~\ref{nolinesfig}).
Note that the mean [median] log$_{10}$ far-IR luminosity of this
subsample, 13.2 [13.1], for an average [median] photometric
redshift of 3.79 [3.70], which is 0.3--0.4\,dex below that of the
sample in which line emission has been detected.  For a {\sc
  fwhm} line width of 500\,\kms\ and typical brightness
temperature and \lir/CO ratios (see later, \S\ref{sk}), this
equates to a peak line flux density in CO(4--3) of 1.6
[1.4]\,mJy, comparable to the r.m.s\ noise levels in our spectral
scans, which goes some considerable way towards explaining why we
detected no line emission for this sub-sample.

\begin{figure}
\centerline{\includegraphics[width=3.5in]{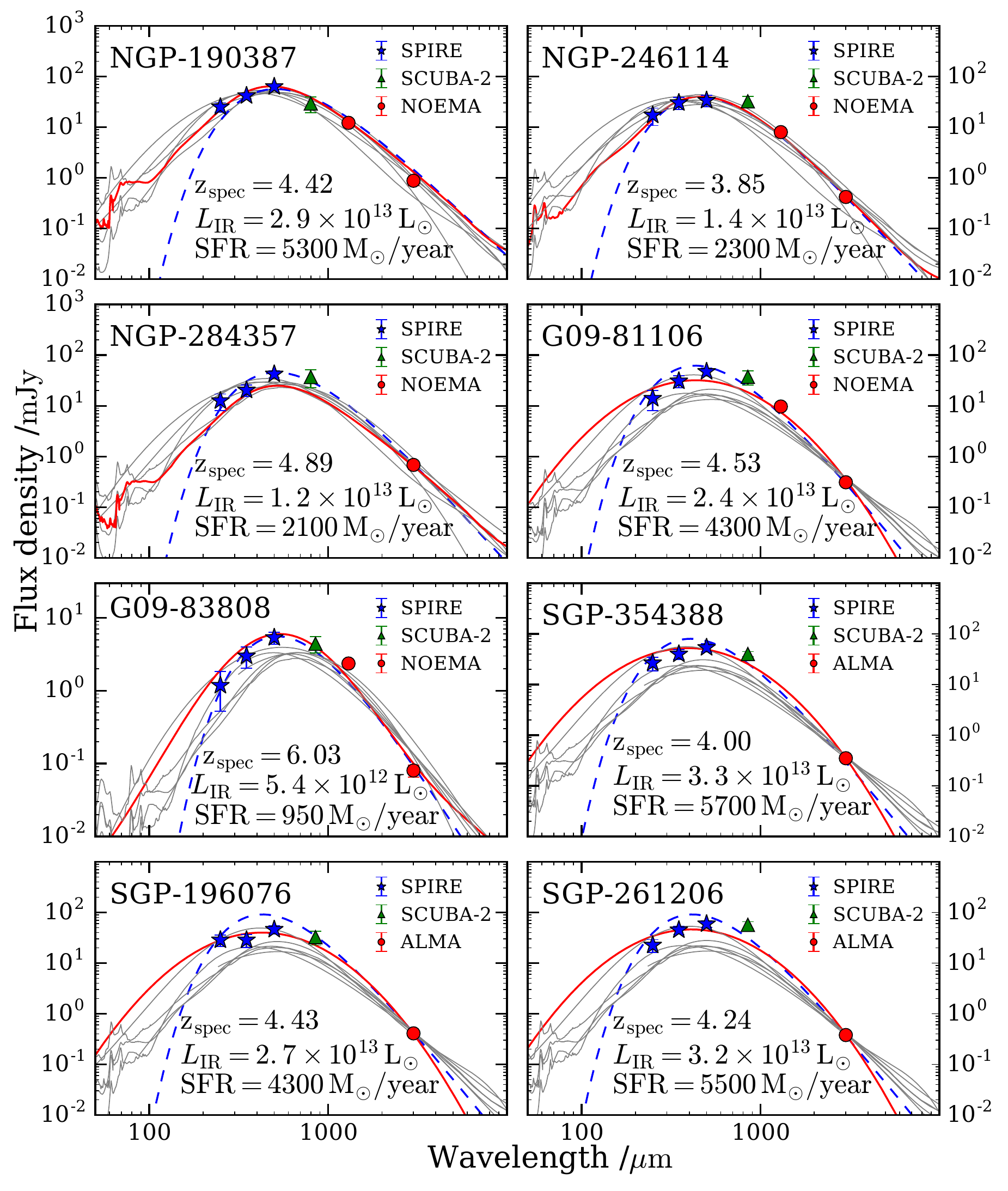}}
\caption{SEDs of those galaxies with unambiguous spectroscopic
  redshifts. Data are from {\it Herschel} SPIRE (blue; 250, 350
  and 500\,$\mu$m), SCUBA-2 (green; 850\,$\mu$m) and NOEMA (red;
  1.3 and/or 3\,mm).  Best-fit SEDs are also shown - red solid
  lines for the best-fit template, blue dashed lines for the
  best-fit modified blackbody function, with other models scaled
  to minimize $\chi^2$ shown as grey solid lines. Best-fit
  templates: \citet{pope08} for NGP-190387; the Cosmic Eyelash
  \citep{swinbank10,ivison10eyelash} for NGP-246114 and
  NGP-284357; G15.141 \citep{cox11} for G09-81106, SGP-354388,
  SGP-196076 and SGP-261206. IR luminosities are calculated using
  the best-fit template between rest-frame 8 and 1000\,$\mu$m,
  and SFR is estimated using these IR luminosities with a
  Salpeter IMF and the empirical calibration of \citet{hao11},
  \citet{murphy11} and \citet{kennicutt12}. The values displayed
  have not been corrected for gravitational amplification, $\mu$
  (see Table~\ref{compare} for $\mu$-corrected values).  Dust
  temperatures and masses determined from the modified blackbody
  fits are listed in Table~\ref{sedfit}.}
\label{sedres}
\end{figure}

\subsection{Spectral energy distributions}
\label{sedfitsec}

Since we have added continuum flux density measurements at 1.3 and/or
3\,mm for many of our targets, as well as some unambiguous
spectroscopic redshifts, it is worth repeating the SED fits performed
by \citet{ivison16}.  We have constructed the SEDs of our targets,
utilizing data from SPIRE at 250, 350 and 500\,$\mu$m, from SCUBA-2 at
850\,$\mu$m \citep{ivison16}, from NOEMA at 1.3\,mm, and from NOEMA
and/or ALMA at 3\,mm -- see Table~\ref{mescont}.  For details of the
SED fits for SGP-32386, we refer readers to \citet{oteo16}.

Like \citeauthor{ivison16}, we employ SED templates representative of
high-redshift DSFGs: the average SEDs from \citet{swinbank14},
\citet{pope08} and \citet{pearson13}, and the observed SEDs of
individual targets -- the Cosmic Eyelash \citep{swinbank10,
  ivison10eyelash}, HFLS\,3 \citep{riechers13}, G15.141 \citep{cox11}
and Arp\,220 \citep{donley07}.

We have restricted our SED work to the sources with unambiguous
redshift determinations, such that we need shift only the flux
density scale of the templates to fit the observed SEDs. We
adopted the lowest $\chi^2$ values, calculated from the
difference between the templates and the observed flux densities,
with inverse weighting of the flux density uncertainties.  These
best-fitting SEDs are plotted in Fig.~\ref{sedres}, with the
corresponding IR luminosities ($L_{\rm 8-1000\,\mu m}$) and SFRs
listed in Table~\ref{compare}, the latter calculated using the
calibration of \citet{hao11}, \citet{murphy11} and
\citet{kennicutt12}, with a Salpeter IMF \citep[although see][for
a cautionary tale regarding the IMF in such
starbursts]{Romano17}.  On the basis of high-resolution ALMA
continuum and line observations, \citet{oteo16} found that
SGP-196076 at $z=4.425$ comprises at least three components,
their on-going merger driving large masses of turbulent gas to
form stars, as is ubiquitous amongst objects with such high
intrinsic IR luminosties \citep[e.g.][]{ivison98, ivison10leblob,
  ivison13, frayer98, frayer99, fu13, bothwell13, messias14,
  rawle14, dye15, thomson15, geach15, oteo16}.

\subsubsection{Modified blackbody fits}

To better quantify the thermal dust emission we have performed SED
fits using modified blackbody (MBB) spectra, again by minimizing
$\chi^2$.  We adopted an optically thin model with single dust
temperature (i.e.\
$S_{\nu}(T_{\rm d})\propto (\frac{\nu}{\nu_{\rm
    c}})^{\beta}\,B_{\nu}(T_{\rm d})$), where $\nu_c$ is the frequency
at which the optical depth is unity, $B_{\nu}(T_{\rm d})$ is Planck
function at frequency, $\nu$, and dust temperature, $T_{\rm d}$, and
$\beta$ is the dust emissivity index.  We fixed $\nu_{\rm c}$ to
1.5\,THz \citep[e.g.][]{conley11,rangwala11} and adopted
$\kappa_{850\,{\rm \mu m}} = 0.15\,{\rm m^2\,kg^{-1}}$
\citep{weingartner01,dunne03}.  Dust emissivity, $\beta$, being poorly
constrained by our data, was fixed to values of 1.5, 2.0 or 2.5
\citep{dunne01,magnelli12,casey11,chapin11,walter12}.

The assumption of a single dust temperature means that we are
measuring the emission-weighted mean dust temperature and dust mass of
all the dust components in the galaxy.  Another advantage of this
approach is that it allows us to compare directly with other
high-redshift DSFGs, which are usually described in terms of
single-temperature MBB fits (see Table~\ref{compare}).  Finally, the
modest sampling of our SEDs, especially at the short wavelengths
required to constrain hot dust components, prevents meaningful
multi-temperature MBB fitting.

The best MBB fits are plotted in Fig.~\ref{sedres}.  Minimum $\chi^2$
were obtained with $\beta=1.5$ for NGP-190387, $\beta=2.0$ for
NGP-246114 and NGP-284357.  G09-81106 proved difficult to reconcile
with a value of $\beta$ below 2.5 -- probably the result of using a
single temperature MBB \citep[see, e.g.,][]{walter12}.  Changing
$\beta$ results in an increase/decrease of the dust temperature by
$\sim 2$--3\,{\sc k}, and the dust mass by $\sim 0.1$\,dex.  The
resulting dust temperatures are in the range $\sim31$--36\,{\sc k},
and dust masses, $\sim 0.2$--$4.1\times10^{9}$\,M$_{\odot}$. For their
high IR luminosities, the dust temperatures of our galaxies are
relatively low \citep[see, e.g.,][]{symeonidis13}.  Individual values
of $T_{\rm d}$ and $M_{\rm d}$ are listed in Table~\ref{sedfit}.

\begin{table}
	\begin{center}
	\caption{Dust temperatures and masses.}
	\label{sedfit}
	\begin{tabular}{lccc} 
		\hline\hline
		Object&$T_{\rm d}$ /K&$M_{\rm d}$ /M$_{\odot}$&$\beta$\\
		\hline
 SGP-196076$^{\rm a} $&$\sim 33$&$\sim2.6\times10^{9}$&2.0\\
 SGP-261206$^{\rm b} $&$33.1\pm0.5$&--&2.5\\
 SGP-354388&$32.3\pm 0.8$&$1.4\pm 0.2\times10^9$&2.5\\
 G09-81106&$34.7\pm1.0$&$8.6\pm1.0\times10^8$&2.5\\
 G09-83808$^{\rm c}$ &$35.9\pm 1.5$&$ 2.1\pm 0.3\times10^8$&2.5\\
 NGP-284357&$33.3\pm1.4$&$2.3\pm0.4\times10^9$&2.0\\
 NGP-190387$^{\rm b}$ &$34.4\pm1.1$&--&1.5\\
 NGP-246114&$30.7\pm1.1$&$2.0\pm0.3\times10^9$&2.0 \\
		\hline
	\end{tabular}
	\end{center}
	$\rm^{a}$ From Table~2 of \citet{oteo16}: $T_{\rm d}$
   is average of brightest two components; dust mass is
   their sum.
	$\rm^{b}$ SGP-261206 and NGP-190387 are gravitationally lensed; only $T_{\rm d}$ is constrained as the magnification factors are unknown.
	 $\rm^{c}$ G09-83808 is gravitationally amplified by a factor 
   $8.2\pm 0.3$ \citep{oteo17morph}.
\end{table}

\begin{table*}
	\begin{center}
	\caption{Properties compared to other $z\gtrsim4$ galaxies. }
	\label{compare}
	\begin{tabular}{lcccccl}
		\hline\hline
Name&Redshift&SFR$^{\rm a}$&$M_{\rm H_2}^{\rm b}$&$L_{\rm IR}$&Known to& Reference\\
&&/M$_{\odot}\,{\rm yr^{-1}}$&/$10^{11}$\,M$_{\odot}$&/$10^{13}$\,L$_{\odot}$&be lensed?&\\
		\hline
SGP-261206&4.242&$5500/\mu$&$3.2/\mu$&$1.3/\mu$&Yes$^{\rm c}$& This work\\
G09-81106&4.531 &$4300$ &$1.2$&$2.4$&No & This work\\
NGP-284357&4.894 &$2100$ &$2.1$&$1.2$&No & This work\\
NGP-190387&4.420 &$5300/\mu$ &$2.1/\mu$&$2.9/\mu$&Yes$^{\rm c}$ & This work \\
NGP-246114&3.847 &$2300$ &$0.99$&$1.4$&No & This work \\
HFLS\,1&4.29&9700&--&$5.6$&No&\citet{dowell14} \\
G09-83808&6.027&$7800/\mu$ &$0.78/\mu$&$4.4/\mu$&Yes$^{\rm c}$ & \citet{zavala17}\\
ADFS-27&5.655&4200&$2.5$&$2.5$&No&\citet{riechers17} \\
SGP-354388&4.002&$5700$ & -- &$3.3$&No & \citet{oteo17grh}\\
SGP-196076&4.425 &$4300^{\rm d}$ &$2.7$&$2.4$ &No & \citet{oteo16}\\
GN\,20&4.055 &$3000$ &$1.3$&$2.9$&No & \citet{hodge12}\\
HDF\,850.1&5.183 &$850/\mu$ &$0.35$&$0.65$ &Weakly$^{\rm c}$ & \citet{walter12}\\
AzTEC-1&4.342&$2000$&$1.4/\mu$&$1.4/\mu$& No& \citet{yun15}\\
AzTEC-3&5.299 &$1800$ &$0.53$&$1.1$&No & \citet{riechers14}\\
HFLS\,3&6.337 &$2100/\mu$ &$0.36/\mu$ &$2.9/\mu$&Weakly$^{\rm c}$ & \citet{riechers13}\\
		\hline
	\end{tabular}
	\end{center}
	\begin{flushleft}
	$\rm^{a}$ SFRs derived from \lir\ (rest-frame
	   8--1000\,$\mu$m), and empirical calibration by \citet{hao11,murphy11,kennicutt12}, i.e.\
	   ${\rm SFR}=L_{\rm IR}/2.21\times10^{43}\,{\rm M_{\odot}\,yr^{-1}}$, adjusting to
	   a Salpeter IMF.\\
	$\rm^{b}$ Molecular gas masses, assuming
	   $\alpha_{\rm CO} = M_{\rm gas}/L^{\prime}_{\rm CO}=0.8\,{\rm
	     K\,km\,s^{-1}\,pc^2}$, typical for high-redshift starbursts and a
	   small sample of local ULIRGs \citep{downes98}.\\
	$\rm^{c}$ For lensed galaxies, listed values are uncorrected for the
	  magnification, $\mu$.\\
	$\rm^{d}$ Total SFR of the most luminous two components \citep{oteo16}.
	\end{flushleft}
\end{table*}

\subsection{Molecular gas masses}

Although low-$J$ transitions of CO, or C\,{\sc i}, are much preferred
when tracing the remaining reservoirs of molecular hydrogen
\citep{ivison11,papadopoulos12}, we can estimate those gas masses from
higher-$J$ CO lines by assuming the average CO line ratio for SMGs, as
measured by \citet{greve05,tacconi06,tacconi08,ivison11,riechers11}
and \citet{bothwell13}, and tabulated by the latter:
$L^{\prime}_{\text{CO(7-6)}}/L^{\prime}_{\text{CO(1-0)}} = 0.18$,
$L^{\prime}_{\text{CO(6-5)}}/L^{\prime}_{\text{CO(1-0)}} = 0.21$,
$L^{\prime}_{\text{CO(5-4)}}/L^{\prime}_{\text{CO(1-0)}} = 0.32$ and
$L^{\prime}_{\text{CO(4-3)}}/L^{\prime}_{\text{CO(1-0)}} = 0.41$,
bearing in mind that there can be large variations.  We find that
$L_{\rm IR}-L^{\prime}_{\rm CO}$ as derived from different -- usually
neighbouring -- transitions are consistent.  We have taken the average
$L^{\prime}_{\text{CO(1-0)}}$ from the available high-$J$ CO
transitions, then calculated the molecular gas masses using
$\alpha_{\rm CO}=0.8\,{\rm M_{\odot}\,(K\,km\,s^{-1}\,pc^{2})^{-1}}$,
the value often assumed for high-redshift starbursts and local ULIRGs
since the work of \citet{downes98}. Estimates of gas-to-dust mass
ratios then lie in the range 50--140, consistent with those of local
galaxies. Our $M_{\rm H_{2}}$ estimates are listed in
Table~\ref{compare}. 

Since luminous CO(5--4) and CO(4--3) emission can be generated by the
presence of a massive molecular gas reservoir, or by a much smaller
amount of highly excited molecular gas, follow-up observations of
low-$J$ CO transitions are required to better determine
$M_{\rm H_{2}}$, modulo the effects of cosmic rays laid out by
\citet{bisbas17}.

\begin{figure}
\centerline{\includegraphics[width=3.5in]{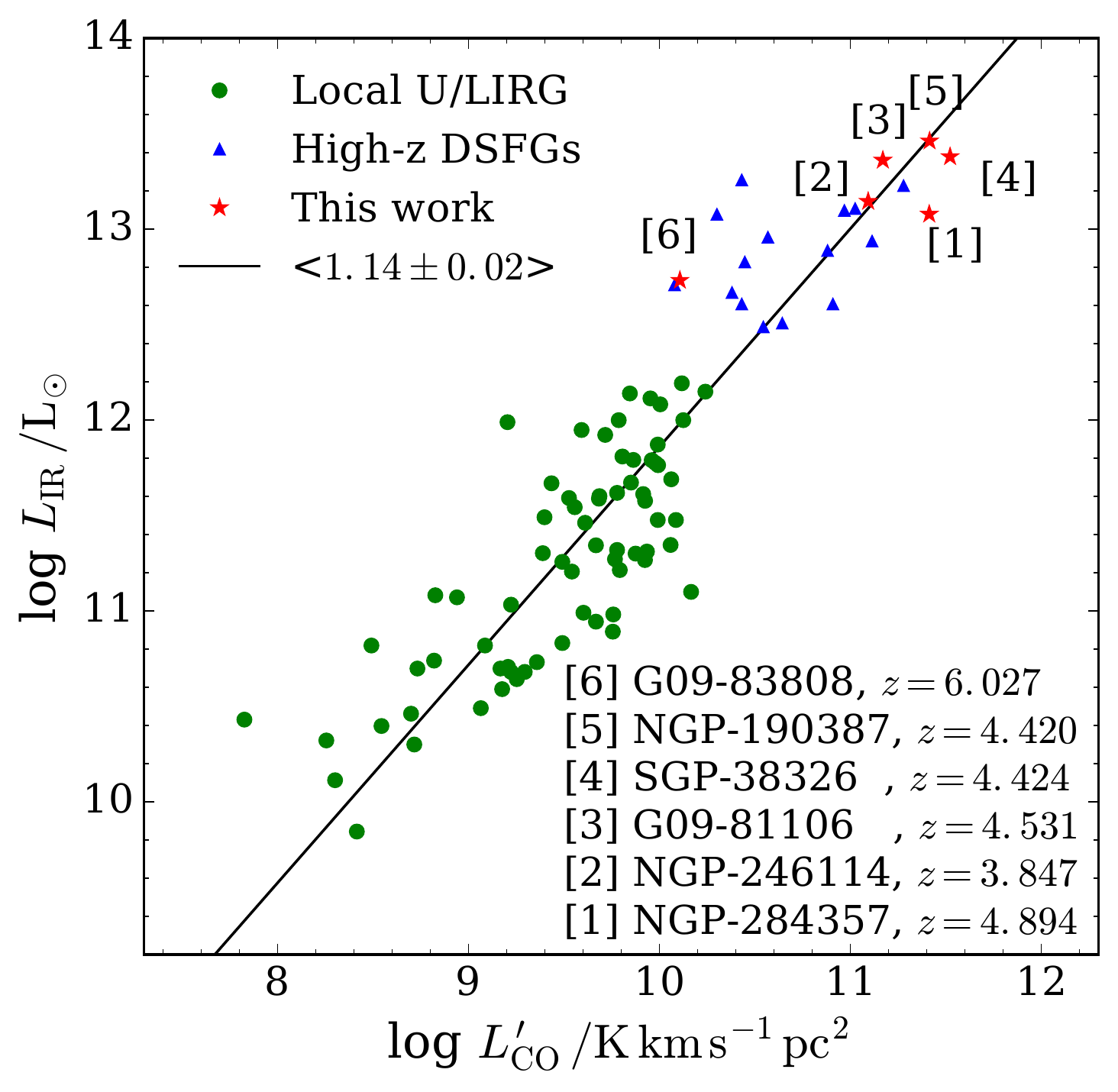}}
 \caption{$L_{\rm IR}-L^{\prime}_{\rm CO}$ correlation of local
   U/LIRGs and high-redshift DSFGs. Green circles: CO(1--0)
   observations of local U/LIRGs \citep{papadopoulos12}; blue
   triangles: high-redshift DSFGs with CO(1--0) observations
   \citep{ivison11} and with mid-$J$ CO observations
   \citep{bothwell13,greve14}.  For SGP-196076 we used the sum of the
   two most luminous components of the merging system. Mid-$J$
   transitions have been converted to CO(1--0) line luminosities using
   line luminosity ratio tabulated by \citet{bothwell13}. Black solid
   line: linear fit to all data points (i.e.\
   ${\rm log}L_{\rm IR} = \alpha\,{\rm log}L^{\prime}_{\rm CO} +
   \beta$, where $\alpha$ and $\beta$ are free parameters).  The
   resulting slope, $\alpha=1.14\pm 0.02$.}
\label{sfe}
\end{figure}

\subsection{$L_{\rm IR} - L^{\prime}_{\rm CO}$ correlation}
\label{sk}

The relationship between star formation and total molecular gas
content is often shown via a plot of the key observables, $L_{\rm IR}$
versus $L^{\prime}_{\rm CO}$, and can reveal if and how star-formation
efficiency (SFE) changes with the amount of molecular gas available
for star formation.  We have constructed a plot of
$L_{\rm IR} - L^{\prime}_{\rm CO}$ using our $z>4$ IR-luminous
galaxies, other high-redshift unlensed DSFGs
\citep{ivison11,bothwell13,greve14} and local U/LIRGS
\citep{papadopoulos12} --- see Fig.~\ref{sfe}.  A linear fit to all
the data has a slope, $1.15\pm0.02$ \citep[see also,
e.g.,][]{iono09,genzel10,ivison11,bothwell13}.

Caution is required here, however, since most of the high-redshift
targets, ours included, are detected in mid-$J$ CO transitions. Using
only CO(1--0) observations, with self-consistent determinations of IR
luminosity, \citet{ivison11} reported a slope significantly below
unity, showing that adopting mid-$J$ CO transitions for high-redshift
galaxies and CO(1--0) transitions for low-redshift galaxies may
artificially steepen the slope.  Differential amplification is likely
also an issue for the lensed galaxies in Fig.~\ref{sfe}, where the
amplifications derived for the dust, the CO(1--0) and/or high-$J$ CO
lines likely differ significantly. Finally, we note that several
studies have suggested that our adopted value of $\alpha_{\rm CO}$ is
too low, including \citet{weiss07} and \citet{papadopoulos12}; indeed,
if we were to apply the formalism of \citet{scoville16}, who use
optically thin long-wavelength dust emission to probe the mass of
molecular gas, we arrive at a value $\sim 3.5\times$ higher, the
equivalent of $\alpha_{\rm CO}\approx 3\,{\rm
  M_{\odot}\,(K\,km\,s^{-1}\,pc^{2})^{-1}}$.

\subsection{Depletion timescale}

The average gas-depletion timescale of our galaxies, $t_{\rm depl}$,
is around 50\,Myr, modulo the possibility of considerably higher gas
masses noted in \S\ref{sk}.  Taken at face value, this $t_{\rm depl}$
is consistent with the idea that our targets are rapidly building a
significant mass of stars, which may be picked up in a later phase at
$z\sim2$--3 as massive `red-and-dead' galaxies by near-IR imaging
surveys \citep[e.g.][]{cimatti04,trujillo06,dokkum10}.

\section{Conclusions}

We report spectral scans of {\it Herschel}-selected ultra-red
galaxies with photometric redshifts estimated to lie at
$\gtrsim4$. For each of 21 galaxies we have covered
$\Delta\nu\approx 20$\,GHz using ALMA and NOEMA in the 3-mm
waveband.  We have determined the redshifts of seven galaxies
unambiguously, in the range $z=3.85$--6.03, detecting multiple
emission lines, usually CO rotational transitions. One of these
redshifts was determined independently by \citet{zavala17}.

For an additional four galaxies, single emission lines are
detected, one of which has been shown by \citet{oteo17grh} to lie
at $z=4.002$. Candidate redshifts are suggested, based on their
photometric redshifts. Follow-up observations are required to
measure their redshifts unambiguously, except in that one case.

Since the comparison of photometric and spectroscopic redshifts
for this sample by \citet{ivison16}, two new spectroscopic
redshifts have been determined, one below and one above the
respective photometric redshifts.  Although the offsets for these
two galaxies are larger than the expected uncertainties in
$z_{\rm phot}$, the overall scatter in
$(z_{\rm phot} - z_{\rm spec}) / (1 + z_{\rm spec})$ is still
consistent with (actually, slightly better than) that of the
training set. In the worst case, the offset can be understood in
terms of contamination of flux densities measured at
$\ge500$\,$\mu$m by a cluster of dusty galaxies
\citep{oteo17grh}.

Our sample of redshift-confirmed galaxies contains extraordinarily
IR-luminous starbursts, with an average SFR of
$\approx 2900$\,M$_{\odot}$\,yr$^{-1}$.  They are also among the most
massive known, in terms of molecular gas mass, and dust mass, with
$M_{\rm H_2}\approx 1.8\times10^{11}\,{\rm M_{\odot}}$ on average, and
$M_{\rm d}\sim 0.9$--$4.1\times10^{9}$\,M$_{\odot}$.

Lurking amongst our IR-luminous galaxies we find three lensed systems.
These would otherwise have been hailed as the most luminous known
starbursts.  It is notable that the vast majority of the brightest
systems selected by {\it Herschel} have been revealed as either lensed
galaxies, groups/clusters of starburst galaxies, starbursts with
buried AGN, or some combination of the three
\citep[e.g.][]{ivison13,oteo16}, which suggests strongly that there
exists a limit to the luminosity of individual starbursting galaxies.

Combining local U/LIRGs, other high-redshifts DSFGs and our new
redshift-confirmed galaxies, the resulting
$L_{\rm IR} - L^{\prime}_{\rm CO}$ correlation has slope close to
unity, $1.14\pm0.02$, suggesting slightly higher
  star-formation efficiency in the most IR-bright galaxies.
 
The gas-depletion timescale of our galaxies, around 50\,Myr, is
consistent with the idea that our targets may be picked up in a later
phase at $z\sim2$--3 as massive `red-and-dead' galaxies by near-IR
imaging surveys.
 
\section*{Acknowledgements}

RJI, IO, VA, LD, SM, JMS and ZYZ acknowledge support from the
European Research Council in the form of the Advanced
Investigator Programme, 321302, COSMICISM.  JMS also acknowledges
financial support through an EACOA fellowship.  DR acknowledges
support from the National Science Foundation under grant number
AST-1614213. HD acknowledges financial support from the Spanish
Ministry of Economy and Competitiveness (MINECO) under the 2014
Ram\'{o}n y Cajal program, MINECO RYC-2014-15686. We thank the
referee, Francoise Combes, for her rapid and constructive
feedback. This work was based on observations carried out with
the IRAM Interferometer, NOEMA, supported by INSU/CNRS (France),
MPG (Germany), and IGN (Spain).  This paper makes use of the
following ALMA data: ADS/JAO.ALMA\#2013.1.00499.S. ALMA is a
partnership of ESO (representing its member states), NSF (USA)
and NINS (Japan), together with NRC (Canada) and NSC and ASIAA
(Taiwan) and KASI (Republic of Korea), in cooperation with the
Republic of Chile.  The Joint ALMA Observatory is operated by
ESO, AUI/NRAO and NAOJ. Based on observations collected at the
European Organisation for Astronomical Research in the Southern
Hemisphere under ESO programme 090.A-0891(A).  Based on
observations obtained at the Gemini Observatory, which is
operated by the Association of Universities for Research in
Astronomy, Inc., under a cooperative agreement with the NSF on
behalf of the Gemini partnership: the National Science Foundation
(United States), the National Research Council (Canada), CONICYT
(Chile), Ministerio de Ciencia, Tecnolog{\'i}a e Innovaci{\'o}n
Productiva (Argentina), and Minist{\'e}rio da Ci{\^e}ncia,
Tecnologia e Inova\c{c}{\~a}o (Brazil).




\bibliographystyle{mnras}
\bibliography{ref}



.


\bsp	
\label{lastpage}
\end{document}